\documentclass[10pt,conference]{IEEEtran}

\usepackage{amsmath,amsfonts,bm}

\def\eqref#1{equation~\ref{#1}}

\def\1{\bm{1}}

\def\vs{{\bm{s}}}

\def\vx{{\bm{x}}}

\def\mS{{\bm{S}}}

\def\mW{{\bm{W}}}
\def\mX{{\bm{X}}}

\DeclareMathAlphabet{\mathsfit}{\encodingdefault}{\sfdefault}{m}{sl}
\SetMathAlphabet{\mathsfit}{bold}{\encodingdefault}{\sfdefault}{bx}{n}

\usepackage{cite}
\usepackage{amsmath,amssymb,amsfonts}
\usepackage{algorithmic}
\usepackage{graphicx}
\usepackage{textcomp}
\usepackage{xcolor}
\usepackage[hyphens]{url}
\usepackage{fancyhdr}
\usepackage{hyperref}
\usepackage{epstopdf}
\usepackage{subfigure}
\usepackage{lineno}
\usepackage{float}
\usepackage{multirow}
\usepackage{multicol}
\usepackage{algorithm}
\usepackage{algorithmic}
\usepackage{array}
\usepackage{xcolor}
\usepackage{booktabs}
\usepackage{color}
\usepackage{soul}
\usepackage{colortbl}
\soulregister{\cite}7 
\soulregister{\citep}7 
\soulregister{\citet}7 
\soulregister{\ref}7 
\soulregister{\pageref}7 
\soulregister{\caption}7 
\definecolor{yellow1}{RGB}{255, 242, 204}
\definecolor{orange1}{RGB}{248, 203, 173}
\definecolor{blue1}{RGB}{180, 199, 231}
\definecolor{green1}{RGB}{226, 240, 217}
\definecolor{purple1}{RGB}{160, 138, 214}
\definecolor{yellow2}{RGB}{255, 220, 0}
\definecolor{orange2}{RGB}{248, 203, 173}
\definecolor{blue2}{RGB}{180, 199, 231}
\definecolor{green2}{RGB}{226, 240, 217}
\definecolor{purple2}{RGB}{175, 33, 151}
\graphicspath{{figures/}}

\newcommand{\hpcayear}{2024}

\newcommand{\hpcasubmissionnumber}{679}
\title{MEGA: A Memory-Efficient GNN Accelerator Exploiting Degree-Aware Mixed-Precision Quantization}

\def\hpcacameraready{} 

\renewcommand{\thefootnote}{\fnsymbol{footnote}}
\newcommand\hpcaauthors{Zeyu Zhu$^{1,2\dagger}$ \
Fanrong Li$^{1\dagger}$ \ Gang Li$^{3}\ddagger$
\ Zejian Liu$^{1}$ \ Zitao Mo$^{1}$ \
Qinghao Hu$^{1}$ \ Xiaoyao Liang$^{3}$ \ Jian Cheng$^{1,2,4,5}\ddagger$}
\newcommand\hpcaaffiliation{
$^1$Institute of Automation, Chinese Academy of Sciences\\
$^2$School of Future Technology, University of Chinese Academy of Sciences\\
$^3$Shanghai Jiao Tong University\\
$^4$AiRiA \ 
$^5$ Maicro.ai\\
}
\newcommand\hpcaemail{\{zhuzeyu2021, 
lifanrong2017, liuzejian2018, huqinghao2014, mozitao2017\}@ia.ac.cn \\
gliaca@sjtu.edu.cn \ liang-xy@cs.sjtu.edu.cn \
jian.cheng@ia.ac.cn
}

\author{
  \ifdefined\hpcacameraready
    \IEEEauthorblockN{\hpcaauthors{}}
      \IEEEauthorblockA{
        \hpcaaffiliation{} \\
        \hpcaemail{}
      }
  \else
    \IEEEauthorblockN{\normalsize{HPCA \hpcayear{} Submission
      \textbf{\#\hpcasubmissionnumber{}}} \\
      \IEEEauthorblockA{
        Confidential Draft \\
        Do NOT Distribute!!
      }
    }
  \fi 
}

\fancypagestyle{camerareadyfirstpage}{%
  \fancyhead{}
  
  \fancyhead[C]{
    \ifdefined\aeopen
    \parbox[][12mm][t]{13.5cm}{\hpcayear{} IEEE International Symposium on High-Performance Computer Architecture (HPCA)}    
    \else
      \ifdefined\aereviewed
      \parbox[][12mm][t]{13.5cm}{\hpcayear{} IEEE International Symposium on High-Performance Computer Architecture (HPCA)}
      \else
      \ifdefined\aereproduced
      \parbox[][12mm][t]{13.5cm}{\hpcayear{} IEEE International Symposium on High-Performance Computer Architecture (HPCA)}
      \else
      \parbox[][0mm][t]{13.5cm}{\hpcayear{} IEEE International Symposium on High-Performance Computer Architecture (HPCA)}
    \fi 
    \fi 
    \fi 
    \ifdefined\aeopen 
      \includegraphics[width=12mm,height=12mm]{ae-badges/open-research-objects.pdf}
    \fi 
    \ifdefined\aereviewed
      \includegraphics[width=12mm,height=12mm]{ae-badges/research-objects-reviewed.pdf}
    \fi 
    \ifdefined\aereproduced
      \includegraphics[width=12mm,height=12mm]{ae-badges/results-reproduced.pdf}
    \fi
  }
  \fancyfoot[C]{}
}
\begin{document}
\maketitle

\ifdefined\hpcacameraready 
  \thispagestyle{camerareadyfirstpage}
  \pagestyle{empty}
\else
  \thispagestyle{plain}
  \pagestyle{plain}
\fi

\newcommand{\hpcaheight}{0mm}
\ifdefined\eaopen
\renewcommand{\hpcaheight}{12mm}
\fi

\begin{abstract}
  Graph Neural Networks (GNNs) are becoming a promising technique in 
  various domains due to their excellent capabilities in 
  modeling non-Euclidean data. 
  Although a spectrum of accelerators has been proposed to accelerate the inference 
  of GNNs, our analysis demonstrates that 
  the latency and energy consumption induced by DRAM access still 
  significantly impedes the improvement of performance and energy efficiency.
  To address this issue, we propose 
  a \textbf{\underline{M}}emory-\textbf{\underline{E}}fficient 
  \textbf{\underline{G}}NN \textbf{\underline{A}}ccelerator (\textbf{MEGA}) 
  through algorithm and hardware co-design in this work. Specifically, 
  \underline{at the algorithm level}, through
  an in-depth analysis of the node property, we observe that
  the data-independent quantization in previous works is not optimal in terms
  of accuracy and memory efficiency.
  This motivates us to
  propose the \textbf{Degree-Aware} 
  mixed-precision quantization method, in which a proper bitwidth is learned and
  allocated to a node according to its in-degree to compress GNNs 
  as much as possible while maintaining accuracy. 
  \underline{At the hardware level}, we employ a 
  heterogeneous architecture design in which the aggregation and combination 
  phases are implemented separately with different dataflows. 
  In order to boost the performance and energy efficiency, we also present an 
  \textbf{Adaptive-Package} format to alleviate the storage overhead 
  caused by the fine-grained bitwidth and diverse sparsity, and a \textbf{Condense-Edge} scheduling 
  method to enhance the data locality and further alleviate the access irregularity 
  induced by the extremely sparse 
  adjacency matrix in the graph.
  We implement our MEGA accelerator in a 28nm technology node.
  Extensive experiments demonstrate that 
  MEGA can achieve an average speedup of 
  $38.3\times$, $7.1\times$, $4.0\times$, $3.6\times$ and 
  $47.6\times$, $7.2\times$, $5.4\times$, $4.5\times$ energy savings 
  over four state-of-the-art GNN accelerators, HyGCN, GCNAX, GROW, and SGCN, respectively, while 
  retaining task accuracy.
\end{abstract}

\section{Introduction}
\renewcommand{\thefootnote}{$\dagger$}
\footnotetext[1]{Co-first authors.}
\renewcommand{\thefootnote}{$\ddagger$}
\footnotetext[4]{Co-corresponding authors.}

Recently, Graph Neural Networks (GNNs) have attracted much attention from both industry
and academia due to their superior 
learning and representing ability for non-Euclidean geometric data. A number of GNNs have 
been widely used 
in practical application 
scenarios, such as social network analysis\cite{lerer2019pytorch, fan2019graph}, 
autonomous driving\cite{weng2020joint}, 
and recommendation systems\cite{jin2020multi,yang2020hagerec}. 
Many of these tasks demand low-latency and energy-efficient inference, especially for 
edge computing like autonomous driving \cite{klimke2022cooperative,weng2020joint}.

As presented in MPNN framework\cite{gilmer2017neural},
the inference process of GNNs
contains two 
phases: aggregation and combination, which exhibits unique computational 
characteristics compared to other DNNs such as CNN \cite{krizhevsky2017imagenet,
szegedy2015going,he2016deep} 
and Transformer \cite{vaswani2017attention,dosovitskiy2020image,kenton2019bert}. 
In the aggregation phase, each node in the graph collects 
information from its neighboring nodes
according to the connections. 
Then, in the combination phase, 
the collected features of a node are updated 
using a combination function such as MLP. 
For the real-world graph, the connections are extremely sparse and unstructured. 
{While the high sparsity can be leveraged for reducing computations in the aggregation 
phase, the complex data structure poses a significant challenge to memory access, 
leading to unexpected latency and energy cost, especially for general processors 
like CPU/GPU \cite{nodehi2018tigr,yan2019alleviating}.} In addition, a real-world 
graph often contains a vast volume of nodes, e.g., the Reddit dataset has 232965 
602-dimensional nodes, which exacerbates the complexity of memory access in the 
combination phase. Therefore, addressing the memory bottlenecks in both phases 
is critical for fast and energy-efficient GNN inference.
\begin{figure}[t]
  \centering
  \subfigure[Execution Cycle]{
      \begin{minipage}[t]{0.45\linewidth}
         \centering
         \includegraphics[scale=0.23]{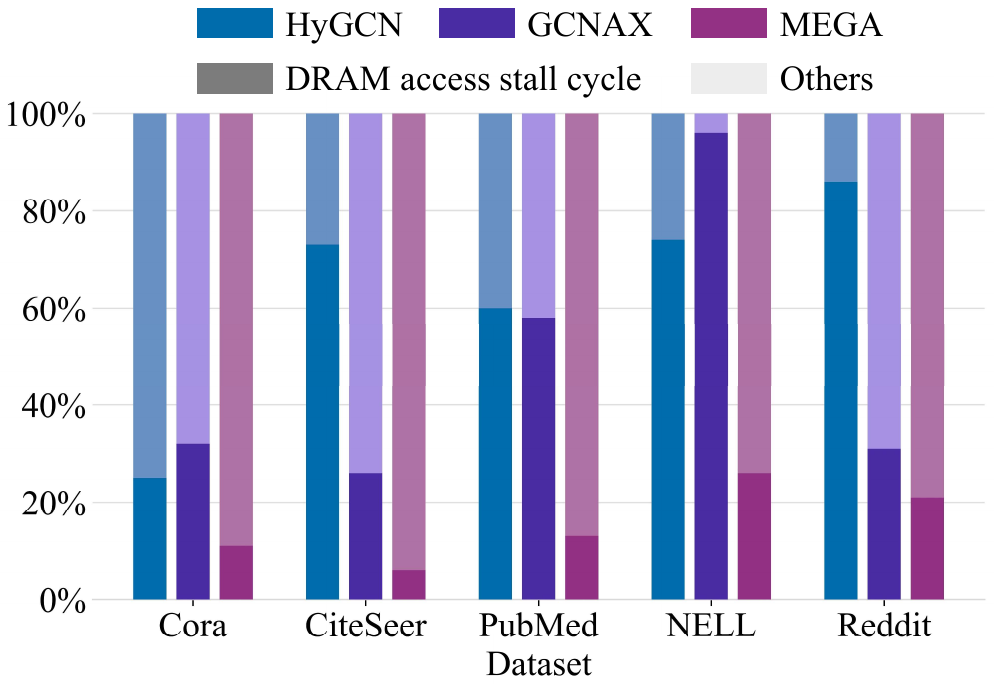}
         \label{hygcn_stall}
      \end{minipage}
  }
   \subfigure[Energy Consumption]{
      \begin{minipage}[t]{0.45\linewidth}
         \centering
         \includegraphics[scale=0.23,trim=0 0cm 0 0, clip]{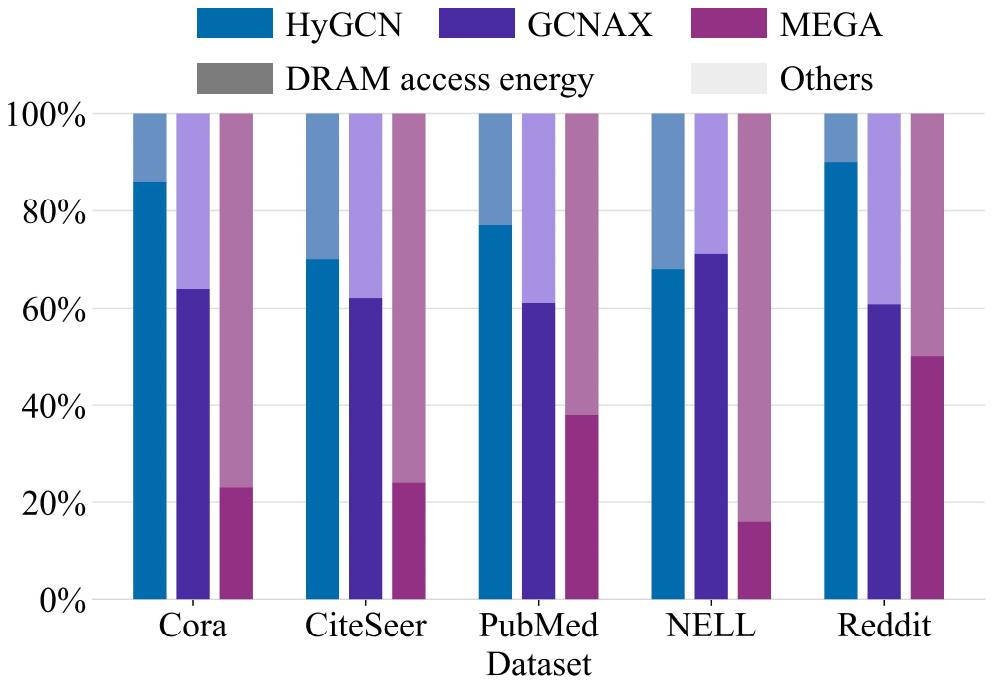}
         \vspace{-0.4cm}
         \label{hygcn_energy}
      \end{minipage}
   }
   \vspace{-0.3cm}
   \caption{The cycle and energy breakdown 
   of HyGCN and GCNAX on different tasks.}
   \vspace{-0.3cm}
   \label{hygcn_analysis}
\end{figure}

Recently, a spectrum of dedicated GNN accelerators has been proposed to improve 
performance and energy-efficiency \cite{yan2020hygcn,li2021gcnax,you2022gcod,hwang2023grow,geng2020awb,liang2020engn,yoo2023sgcn
,zeng2020graphact,chen2021dygnn,chen2022regnn,lin2022hp,sarkar2023flowgnn}.
HyGCN \cite{yan2020hygcn} is one of the first domain-specific accelerators 
for GNNs, which 
designs separate engines for 
aggregation and combination, respectively, to pipeline the two phases of the forward pass. 
GCNAX \cite{li2021gcnax}
models the execution cycle and DRAM
access according to the loop optimization and then explore the design space by 
enumeration to find an optimal execution method. 
Although these works have optimized the GNN inference at the software and hardware levels, our analysis 
reveals 
that the performance and energy bottleneck caused by DRAM access is still 
very serious. 
As shown in Fig. \ref{hygcn_analysis}, the stall 
caused by DRAM access accounts for up to 86.2\% of the overall execution cycles on HyGCN.
Worse still, the energy 
consumption of DRAM access also dominates the overall energy consumption, 
e.g., 90.2\% for HyGCN on Reddit.

As a promising technique to achieve memory-efficient inference, network quantization that
converts floating-point values into low-precision 
fixed-point counterparts can effectively reduce memory footprint and bandwidth requirement.
Unfortunately, current methods for GNN quantization, such as DQ\cite{tailor2020degree} and 
LP\cite{zhao2020learned}, 
are only capable of quantizing the model to 8bit without accuracy loss. 
If the compression ratio is further increased, such as quantized to 4bit, the model  
accuracy will significantly decline.
Through an in-depth analysis 
of the graph nodes, we note that the uniform quantization 
(a specific quantization bitwidth is shared among all nodes) 
in existing works fails to distinguish the importance of the nodes. 
In this scenario, the nodes that are more sensitive to quantization 
will impede the overall bitwidth reduction. Besides, allocating the 
same bitwidth to the less important nodes will lead to considerable memory costs. 
Therefore, taking advantage of the node characteristics for quantization is 
essential to achieve a higher compression ratio while maintaining accuracy. 

To unleash the potential of GNN acceleration, in this paper, we propose 
a \textbf{\underline{M}}emory-\textbf{\underline{E}}fficient 
\textbf{\underline{G}}NN \textbf{\underline{A}}ccelerator (\textbf{MEGA}) 
through algorithm and hardware co-design. At the algorithm level, 
we develop a Degree-Aware mixed-precision quantization method for GNN 
compression. Our key observation is that nodes with higher 
in-degrees, which we refer to as important nodes, 
tend to have larger 
average values of features after aggregation, and a smaller fraction of nodes in a graph 
are deemed important. This motivates a mixed-precision 
quantization method, in which a proper bitwidth is learned and allocated to a 
node according to its in-degree to compress the network as much as 
possible while retaining accuracy. At the hardware level, we employ a 
heterogeneous architecture design in which the aggregation and combination 
phases are implemented separately with different dataflows. In particular, 
bit-serial computation is exploited in the combination phase for precision-scalable 
inference. {We also present an Adaptive-Package method to alleviate the 
storage overhead 
induced by the fine-grained bitwidth and diverse sparsity,} and a Condense-Edge scheduling 
method to enhance the data locality and further reduce DRAM accesses. 

\begin{figure}[t]
  \centering
  \vspace{0cm}
  \includegraphics[scale=0.5,trim=0cm 0cm 0cm 0.2cm,clip]{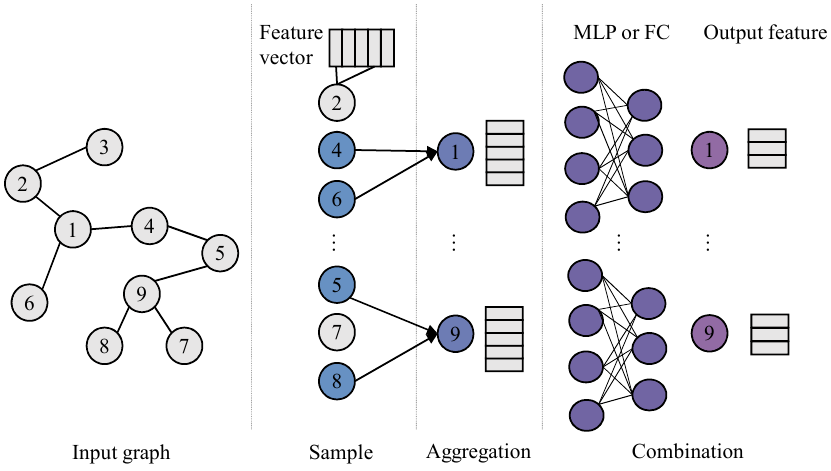}
  \vspace{-0.2cm}
  \caption{Illustration of the GNN models.}
  \vspace{-0.5cm}
\label{aggre_combi_phase}
\end{figure}

In summary, the contributions of this paper are as follows:
\begin{itemize}
  \item We 
  give an in-depth analysis of the limitations in previous GNNs quantization methods 
  and then
  propose the \textbf{Degree-Aware} mixed-precision quantization
  method to enable 
  adaptive learning of 
  quantization parameters for 
  nodes with different degrees, which can improve 
  the model compression ratio as much as possible while maintaining the model accuracy.
  \item We propose an accelerator named \textbf{MEGA} co-designed with our Degree-Aware
  mixed-precision quantization method. It employs a 
  heterogeneous architecture that computes the 
  combination and aggregation phases separately using different dataflows.
  To mitigate the complexity caused by the mixed-precision node features under diverse sparsity,  
  we propose an \textbf{Adaptive-Package} format to improve the efficiency of storage and memory access.
  Moreover, we
  design a \textbf{Condense-Edge} scheduling strategy in MEGA to further reduce the 
  DRAM access.
  \item The experimental results demonstrate that our Degree-Aware quantization can achieve up to 
  15.8\% improvement in accuracy along with a larger compression ratio 
  (up to $18.6\times$
  v.s. 
  $8\times$) 
  compared to the 
  state-of-the-art 
  quantization method \cite{tailor2020degree}. 
  We also implement MEGA on a 28nm technology node and evaluate on five real-world datasets and 
  three GNN models. On average, our MEGA can achieve 
  $38.3\times$, $7.1\times$, $4.0\times$, $3.6\times$ speedups and 
  $47.6\times$, $7.2\times$, $5.4\times$, $4.5\times$ energy savings 
  over HyGCN\cite{yan2020hygcn}, 
  GCNAX\cite{li2021gcnax}, GROW\cite{hwang2023grow}, and SGCN\cite{yoo2023sgcn}, respectively.
\end{itemize} 

\section{Background and Related Work}
\subsection{Graph Neural Networks}
\vspace{-0.1cm}

Most GNNs can be presented using the MPNN framework\cite{gilmer2017neural}, 
in which the forward pass in each layer contains 
two phases as shown in Fig. \ref{aggre_combi_phase}. 
First,  
each node collects information from 
neighboring nodes and uses the aggregation function 
to generate hidden features in the aggregation phase. Second, in the combination phase, 
the hidden features are transformed
into new features by the combination function. 
In order to reduce the computational complexity, sometimes the neighboring nodes 
are sampled before aggregating, then only the sampled 
neighbors are used in the aggregation phase. 
In this paper, we focus on the inference acceleration of three typical GNN models, i.e., 
Graph Convolution Network (GCN)\cite{kipf2016semi},  
Graph Isomorphism Network (GIN)\cite{xu2018powerful}, 
and GraphSage\cite{hamilton2017inductive}. 

The forward pass
of the three models can be uniformly represented as follows:
\setlength{\abovedisplayskip}{3pt}
\setlength{\belowdisplayskip}{3pt}
\begin{equation}
  {\mathbf{X}^{( l )}}
  =\sigma ( \tilde{\mathbf{A}}{\mathbf{X}^{( l-1 )}}{\mathbf{W}^{( l-1 )}} )\text{,}
\end{equation}
where $\tilde{\mathbf{A}}$ is the normalized adjacency matrix, 
$\mathbf{X}^{( l-1 )}$ and $\mathbf{W}^{( l-1 )}$ 
are the feature map including all
nodes features and the learned parameters of the $(l-1)$-th layer, respectively, 
and the $\sigma$ is an activation function, e.g., ReLU. 

For the computation of $\tilde{\mathbf{A}}\mathbf{X}\mathbf{W}$, 
there are two execution patterns, $(\tilde{\mathbf{A}}\mathbf{X})\mathbf{W}$
and $\tilde{\mathbf{A}}(\mathbf{X}\mathbf{W})$. 
We choose $\tilde{\mathbf{A}}(\mathbf{X}\mathbf{W})$ in our accelerator design, 
{which 
greatly reduces the
number of MACs for the GNNs of our focus that have two or three layers and 
the dimension of input features is much larger than the hidden features, }
as reported in \cite{geng2020awb,hwang2023grow,li2021gcnax}.

\vspace{-0.2cm}
\subsection{Quantization}
\vspace{-0.1cm}
Although
some works have applied the quantization technique to GNNs,
there remain some issues that hinder the release of the potential 
to accelerate the inference and 
reduce the memory requirement. 
SGQuant \cite{feng2020sgquant} only quantizes the nodes features and still 
employs floating-point computation
during inference, which is limited in improving the energy efficiency\cite{han2016eie}. 
DQ \cite{tailor2020degree} proposes a degree-quant training strategy, 
which generates masks for nodes with high degrees and only quantizes the nodes without
masks. 
This method can 
quantize GNNs to 8bit with negligible accuracy degradation. 
However, when quantizing
GNNs to lower bitwidth, the accuracy of DQ drops severely. 
There are also some works\cite{wang2021bi,bahri2021binary,wang2021binarized,jing2021meta} that
quantize GNNs to 1-bit.
Though the compression ratio is high, 
these methods will cause unacceptable accuracy degradation on complex GNN models, e.g., GIN and GraphSage, 
which obstructs their generalization.

Moreover, 
based on the idea that different layers have different sensitivities to quantization, 
mixed-precision quantization is proposed in CNNs 
to quantize different layers with different bitwidths
\cite{uhlich2019mixed,esser2020learned,
jain2020trained,dong2019hawq,dong2020hawq,chen2021towards,dai2021vs}.
However, due to the vast differences between GNNs and CNNs, e.g., the extremely sparse 
connections among nodes, 
it is difficult to use these methods on GNNs directly.

\vspace{-0.2cm}
\subsection{Hardware Acceleration of GNNs}
\vspace{-0.1cm}
There are many previous works
\cite{yan2020hygcn,li2021gcnax,you2022gcod,hwang2023grow,geng2020awb,liang2020engn,yoo2023sgcn
,zeng2020graphact,chen2021dygnn,chen2022regnn,lin2022hp,sarkar2023flowgnn} that design dedicated hardware accelerators for GNNs. 
HyGCN\cite{yan2020hygcn} is one of the first domain accelerators that proposes
the window sliding method to eliminate zeros. 
{However, 
ignoring the sparsity existing in nodes features results in low utilization 
of its process engines 
and severe pipeline stalls.}
GCNAX\cite{li2021gcnax} models 
the execution cycle and DRAM access according to the 
loop tile and explores the design space by
enumeration to find 
the optimal tiling pattern.
However, 
GCNAX cannot handle the irregular memory access 
caused by the extremely sparse adjacency matrix. 
GCoD\cite{you2022gcod}
proposes a co-design framework, which can prune, cluster
the adjacency matrix at the algorithm level by using the graph partition algorithm, 
and separately process the dense region
and sparse region through dedicated hardware design.
GROW\cite{hwang2023grow}, which adopts the row product to perform the two phases of GNNs,
also employs 
the same graph partition algorithm to improve the locality of its dataflow.
However, the more sparse connections among different subgraphs 
will lead to severe DRAM accesses and
hinder the improvement of performance
and energy efficiency. 
SGCN\cite{yoo2023sgcn} employs a feature compression format to reduce 
off-chip traffic and designs a dedicated accelerator to 
process the compressed features efficiently. {However, adopting a
systolic array 
to perform the combination phase 
results in SGCN not being able to exploit the sparsity in the combination phase.}

\begin{table}[t]
  \vspace{0cm}
    \caption{The accuracies and compression ratios of GIN when 
    using DQ with different quantization bitwidths
    trained on CiteSeer (CR: Compression Ratio, Acc: Accuracy).}
     \label{dq_quantize_comparison}
    \vspace{-0.4cm}
    \begin{center}
      \begin{tabular}{cccc}
        \hline
            & FP32       & 8bit       & 7bit       \\ \hline
        Acc & 66.1±0.9\% & 67.5±1.4\% & 65.0±2.4\% \\
        CR  & 1x         & 4x         & 4.6x       \\ \hline
            & 6bit       & 5bit       & 4bit       \\ \hline
        Acc & 63.8±3.3\% & 63.7±3.4\% & 60.8±2.1\% \\
        CR  & 5.3x       & 6.4x       & 8x         \\ \hline
        \end{tabular}
    \end{center}
    \vspace{-0.4cm}
\end{table}

\begin{figure}[t]
\vspace{0cm}
\centering
  \includegraphics[scale=0.35]{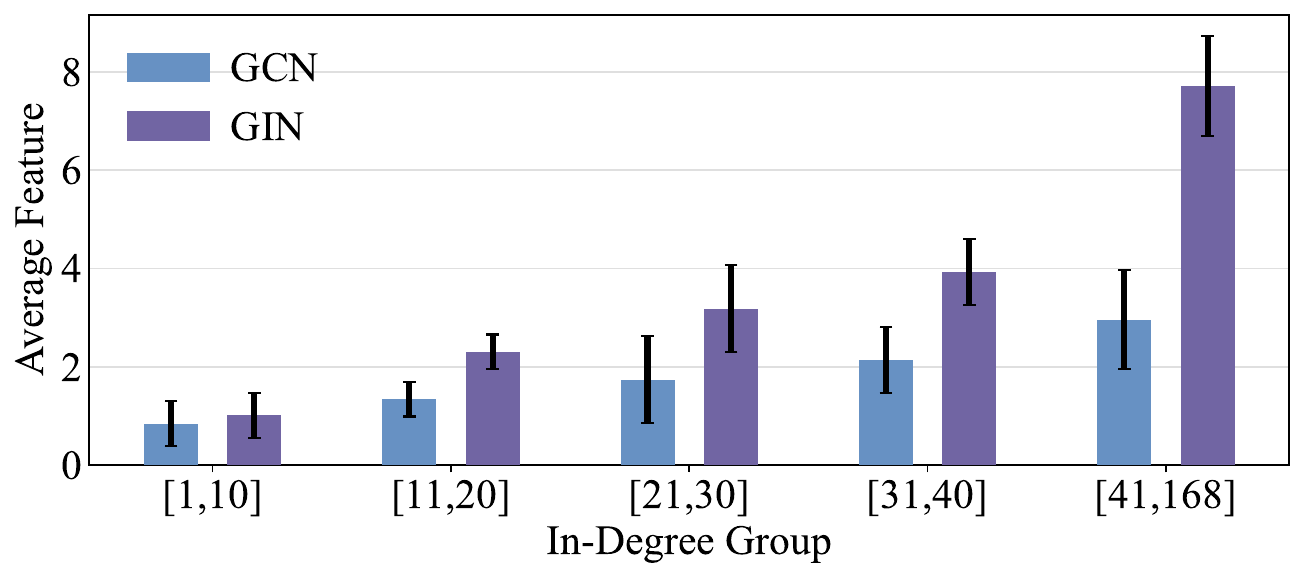}
\vspace{-0.2cm}
\caption{Average node feature values with respect to different in-degrees after aggregation. 
The average values are all generated from 100 runs.
}
\vspace{-0.3cm}
\label{fea_var_deg}
\end{figure}

\begin{figure*}[t]
  \vspace{-0cm}
  \centering
  \includegraphics[scale=0.58]{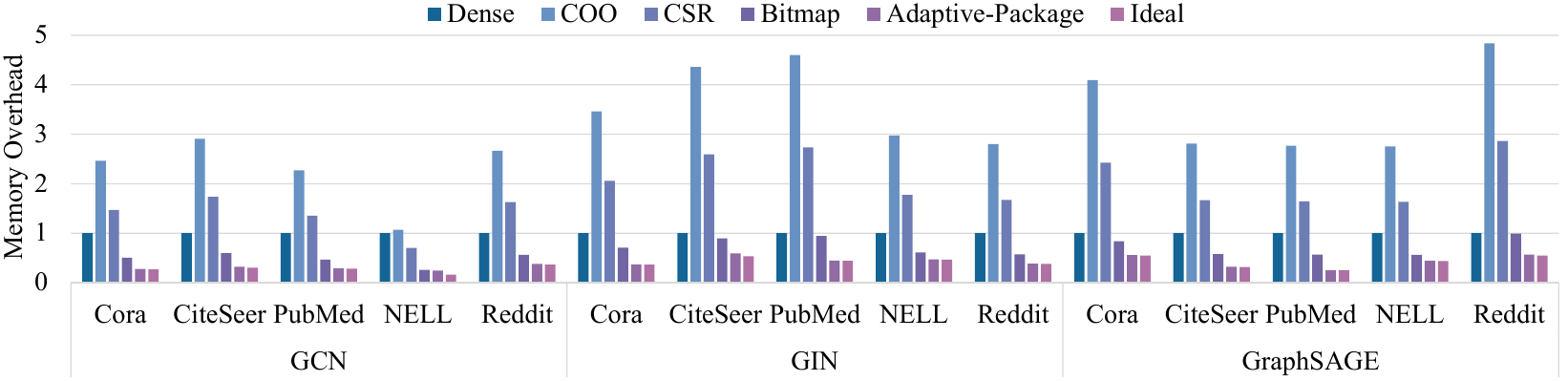}
\vspace{-0.2cm}
\caption{Comparisons of various sparse data representations on three models (normalized to Dense).
Ideal: only quantized non-zero values 
are stored.
Adaptive-Package is our proposed method.}
\vspace{-0.4cm}
\label{sparse_method_cmp}
\end{figure*}
\begin{figure}[t]
  \centering
  \includegraphics[scale=0.65]{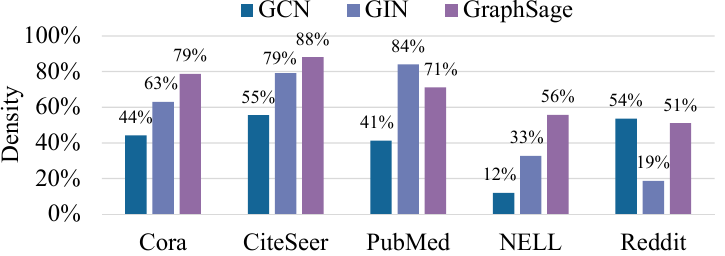}
\vspace{-0.2cm}
\caption{Comparison of sparsity across various datasets on different models.}
\vspace{-0.2cm}
\label{sparsity_various}
\end{figure}

\vspace{-0cm}
\section{Motivation}
\subsection{Rethinking GNN Quantization}
Despite the promising benefits of network quantization in achieving memory-efficient 
inference, prior works have overlooked the potential of utilizing quantization for GNN 
accelerator design. 
{Although there are some works on GNN quantization,
we observe that these existing methods still have limitations. For example, 
the state-of-the-art method DQ\cite{tailor2020degree} can compress the model from 
FP32 to 8bit with negligible accuracy loss. When the quantization bitwidth decreases, 
the accuracy degradation will rise up to 5.3\% (4bit), as shown in 
TABLE \ref{dq_quantize_comparison}.}
Considering the vast amount of DRAM accesses involved in GNN inference, achieving a 
higher compression ratio while maintaining accuracy is crucial to improving 
performance and energy efficiency. 

Note that in existing GNN quantization methods \cite{tailor2020degree,feng2020sgquant}, 
a specific quantization bitwidth 
is shared among all nodes in a graph without considering the graph topology. Through an 
in-depth analysis of the node property, we observe that the data-independent quantization 
is not optimal in terms of accuracy and energy efficiency. Fig. \ref{fea_var_deg} 
visualizes the average aggregated value of node features with respect to different 
in-degrees for GCN \cite{kipf2016semi} and GIN \cite{xu2018powerful} on the Cora dataset. 
It can be noticed that as the in-degree of a node increases, the feature value gets larger 
after aggregation. This indicates that nodes with different in-degrees have different ranges 
of feature values, thus a uniform quantization bitwidth cannot fit the data distribution well. 
Moreover, according to \cite{xie2014distributed,aiello2001random}, 
the in-degrees of nodes in most real-world graphs often follow the power-law distribution, 
i.e., nodes with a low in-degree account for the majority of graph data. Obviously, a 
uniform quantization bitwidth (such as 8bit) for all nodes will cause a significant 
waste on storage and memory accesses. 

Intuitively, to achieve a high compression ratio
while retaining accuracy, 
we can adopt different bitwidths to quantize the nodes with varying degrees, 
since the feature values of a node are closely related to its degree. 
However, due to the 
vast volume of nodes in real-world graphs, manually assigning bitwidths to different nodes 
will result in model accuracy degradation and cannot compress the models to the maximum
extent.
Accordingly, we propose the Degree-Aware mixed-precision 
quantization method in which a proper quantization bitwidth and the associated scale 
are automatically learned and assigned to each node in a graph. 
{To the best of 
our knowledge, we are the first to introduce mixed-precision quantization to GNN 
accelerator design.}

\vspace{-0.2cm}
\subsection{Challenges on Accelerator Design}
\label{challenges on accelerator design}
\vspace{-0.1cm}
Although there have been considerable CNN accelerators 
that support precision-scalable\cite{judd2016stripes,sharma2018bit,sharify2019laconic} 
or sparse computation\cite{han2016eie,zhang2020sparch,kim2017novel,zhang2016cambricon}, 
these accelerators cannot be used directly for GNN acceleration due to the unique 
computation and sparse patterns inherent in GNNs, 
as previously reported\cite{abou2006multilevel,gonzalez2012powergraph,xie2014distributed}.
Furthermore, none of the existing GNN accelerators support mixed-precision data format 
in both computation and storage, impeding the acceleration potentials of our 
Degree-Aware method. Therefore, a dedicated accelerator
co-designed with our proposed algorithm is
essential. However, challenges still remain:

\textbf{\textit{1) Challenge on storage:}}
Properly storing the fine-grained mixed-precision features is critical to
achieving optimal efficiency in GNN accelerator design. However, two 
challenges on storage hinder the gains from highly compressed features.
First,  
the existing sparse data representations 
cannot handle the fine-grained bitwidth 
so the highest quantization bitwidth among all nodes should be used when storing 
the quantized features, leading to non-negligible storage underutilization.
As shown in Fig. \ref{sparse_method_cmp}, 
existing compression formats are far from ideal in the mixed-precision scenario.
Second, identifying the location of non-zero values is challenging.
As shown in Fig. \ref{sparsity_various}, 
the sparsity of nodes features, i.e., $\mathbf{X}$, 
varies greatly across different tasks.
Along with the fine-grained bitwidth, the various sparsity will lead to 
more irregular and random access to the memory and significant overhead on decoding 
the quantized features.

\begin{figure}[t]
  \centering
  \vspace{0cm}
  \includegraphics[scale=0.41,trim=0cm 0.5cm 0cm 0cm, clip]{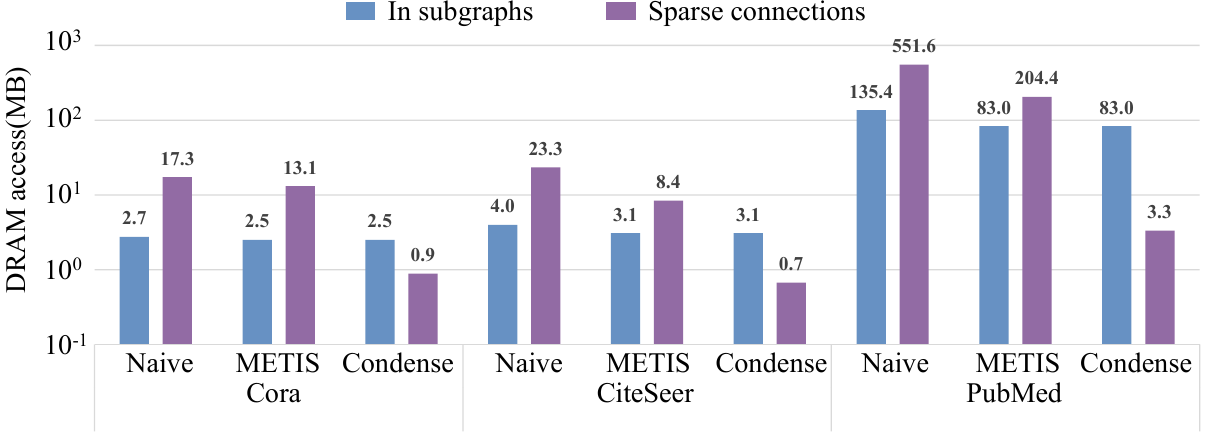}
  \vspace{-0.2cm}
\caption{Comparisons of DRAM access on three configs. Naive: 
without graph partition.
METIS: divide the graph by METIS. 
Condense: our Condense-Edge method. 
}
\vspace{-0.4cm}
\label{dram_condense_cmp}
\end{figure}

\textbf{\textit{2) Challenge on irregular memory access:}}
A great deal of irregular memory accesses to nodes features introduced by the extremely sparse 
connections result in a significant 
discount of the improvement in 
performance and energy savings.
In previous works, 
such as GROW\cite{hwang2023grow} and GCoD\cite{you2022gcod}, 
the graph partition algorithm is used to 
divide the graph into several subgraphs, in which the data locality is significantly improved.
However, this results in more sparse connections between 
subgraphs (referred to as sparse connections), as the total number of edges in the graph 
remains unchanged.
The sparse connections weaken the spatial locality and significantly decrease the data reuse, 
causing more off-chip memory traffic. 
Taking GROW\cite{hwang2023grow} as an example,
we compare the number of DRAM accesses
induced by the dense subgraphs and the 
sparse connections in the aggregation phase with and 
without using the graph partition algorithm METIS\cite{karypis1998fast} 
in Fig. \ref{dram_condense_cmp}. 
We can notice that although the overall 
DRAM accesses can be reduced after graph partition, there are still considerable DRAM accesses
caused by sparse connections. 

In this paper, we present two techniques to address the above issues. Specifically, 
we propose a memory-efficient Adaptive-Package format that can efficiently 
encode/decode the mixed-precision sparse node features and achieve near-ideal 
storage utilization (as shown in Fig. \ref{sparse_method_cmp}). 
We also propose a Condense-Edge scheduling method capable of significantly reducing 
the overhead caused by sparse connections (as shown in Fig. \ref{dram_condense_cmp}).
We will elaborate these techniques in Section~\ref{adaptive_package_section} 
and \ref{condense_edge_section}.

\section{Degree-Aware Quantization}
In this section, we present our Degree-Aware mixed-precision quantization approach, 
which quantizes
nodes with different in-degrees by different learnable quantization parameters, 
including bitwidth and
scale. These parameters can be learned during training. 
By imposing a penalty on memory size, our approach is memory-aware, 
which allows us to obtain a better trade-off between memory footprint and model accuracy. 

\begin{figure}[t]
  \vspace{-0.2cm}
  \flushleft 
  \subfigure[]{
      \begin{minipage}[t]{0.45\linewidth}
         \centering
         \includegraphics[trim=8.5cm 8.5cm 8.5cm 7cm,clip,scale=0.4]{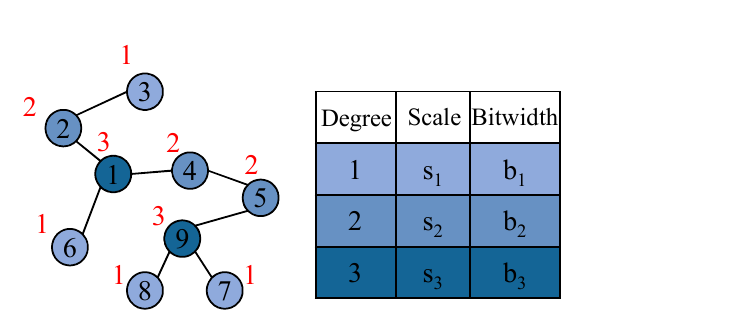}
         \label{quantize}
   \vspace{-0.2cm}
  \end{minipage}
   }
   \subfigure[]{
      \begin{minipage}[t]{0.48\linewidth}
         \centering
         \includegraphics[scale=0.31,trim=8cm 7.5cm 8.5cm 6cm,clip]{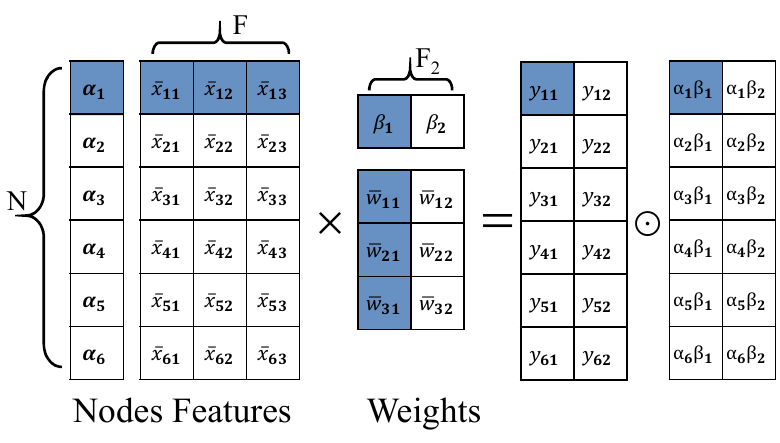}
         \label{matrix_mult}
   \vspace{-0.2cm}
  \end{minipage}
   }
   \vspace{-0.4cm}
   \caption{(a) Quantize different nodes according to their degrees. 
   (b) Perform matrix multiplication by the
   integer represented. $\bar{x}$ and $\bar{w}$ are both integers.}
   \vspace{-0.4cm}
   \label{quantization_method}
\end{figure}

Given a graph with $N$ nodes, we assume that the node features are  $F$-dimensional, 
then the feature map is denoted as $\mathbf{X}\in \mathbb{R}^{N\times F}$, and 
$\vx_i$ is the features of node $i$.
Firstly, we initialize the learnable quantization parameters, i.e., scale 
$\vs=(s_1,s_2,…,s_d)$, and bitwidth $\mathbf{b}=(b_1,b_2,…,b_d)$, where $d$ 
is the maximum degree of nodes
in a graph. We quantize nodes with different degrees as follows:
\setlength{\abovedisplayskip}{3pt}
\setlength{\belowdisplayskip}{3pt}
\begin{gather}
  \bar{\vx}_i=sign(\vx_i) \begin{cases}
     \lfloor \frac{\left| \vx_i \right|}{\alpha_i}+0.5 \rfloor \text{,} \left| \vx \right|<\alpha_i(2^{[b_i]-1}-1) \\
     \\
     2^{[b_i]-1}-1 \text{,} \left| \vx_i \right| \geq \alpha_i(2^{[b_i]-1}-1)  
  \label{quantize_1}
  \end{cases}
  \ \text{,} 
\end{gather}
where $(\alpha_i, b_i)$ are the quantization parameters selected based on 
the degree of the node,
i.e., $\alpha_i = s_{d_i}$ and $b_i=b_{d_i}$, $d_i$ is the degree of the $i$-th node.
Then we can obtain the fixed-point feature map $\bar{\mathbf{X}}$, 
and the original feature can be represented as $\mX_q=\mS_X\cdot \bar{\mathbf{X}}$, 
where $\mS_X=diag(\alpha_1,\alpha_2,...,\alpha_N)$. 
Fig. \ref{quantize} displays the Degree-Aware quantization process.

In the combination phase, the core calculation 
is $\mathbf{X}\mathbf{W}$, where $\mathbf{X}$ 
{is the feature map obtained from the aggregation phase of the previous layer.}
In order to further decrease the redundant DRAM accesses, we also quantize $\mathbf{W}$. 
Given that $\mathbf{W}$ in a specific layer is shared among
all nodes,  
we quantize $\mathbf{W}$ to the same bitwidth of 4bits for all GNNs in this paper. 
Moreover, to enhance the generalization, each column of $\mathbf{W}$ is 
endowed with its individual learnable quantization scale,
i.e.,  $\vs_W=(\beta_1,\beta_2,..,\beta_{F_2})$, where $F_2$ is the 
output-dimension of the node features in the current layer 
and $\beta_i$ is the quantization scale for the $i$-th column of $\mathbf{W}$.
We can obtain the integer representation $\bar{\mathbf{W}}$ as Eq \ref{quantize_1}
and the quantized representation $\mW_q=\bar{\mathbf{W}}\cdot \mS_W$,
where $\mS_W=diag(\beta_1,\beta_2,...,\beta_{F_2})$.  
The float-point matrix multiplication in the combination phase can be reformulated as follow:
\begin{equation}
\setlength{\abovedisplayskip}{3pt}
  \begin{aligned}
    \label{mat_mul}
  \mathbf{X}\cdot \mathbf{W}\approx \mX_q\cdot \mW_q &= (\mS_X\cdot \bar{\mathbf{X}})\cdot (\bar{\mathbf{W}}\cdot \mS_W) \\
                            &= (\bar{\mathbf{X}}\cdot \bar{\mathbf{W}}) \odot (\vs_X \otimes \vs_W)\ \text{,}
  \end{aligned}
  \setlength{\belowdisplayskip}{3pt}
\end{equation}
where $\odot$ denotes an element-wise multiplication, and $\otimes$ denotes the outer product. 
An execution example is illustrated in Fig. \ref{matrix_mult}.
For the aggregation phase, i.e., $\mathbf{A}\mathbf{B}$, 
where $\mathbf{B}=\mathbf{XW}$ and $\mathbf{A}\in\{0,1\}^{N\times N}$ 
is the adjacency matrix,
we quantize $\mathbf{B}$ as the quantization way of $\mathbf{W}$.
Then the aggregation
phase can also be performed by integer operations to reduce the computational overhead.

After quantization, 
the inference process of GNNs can be executed using fixed-point 
operations, 
resulting in a significant reduction in energy consumption for arithmetic operations, 
and a noteworthy decrease in DRAM access.

To better trade off the model accuracy and the memory reduction, 
our Degree-Aware method introduces a penalty on memory size 
to the loss function:
\begin{equation}
  \label{memory_penalty}
\setlength{\abovedisplayskip}{3pt}
{{L}_{memory}}={{( \frac{1}{\eta }\cdot \sum\limits_{l=1}^{L}{\sum\limits_{i=1}^{N}{\dim^{l}\cdot b_{i}^{l}}}-{{M}_{target}})^{2}}} \ \text{,}
\setlength{\belowdisplayskip}{3pt}
\end{equation}
where $L$ is the number of layers in the GNNs, $N$ is the total number of nodes, $\dim^l$ is the 
length of the node features in $l$-th layer,
$b_i^{l}$ is the quantization bitwidth for node $i$ in $l$-th layer, $M_{target}$ is the 
target memory size 
on the total node features memory size,
and $\eta = 8*1024$, which is a constant to convert the unit of memory size to $\rm KB$. 
Then the model and quantization parameters can be trained by the loss function:
\begin{equation}
\setlength{\abovedisplayskip}{3pt}
L_{total}={{L}_{task}}+\lambda \cdot {L}_{memory} \ \text{,}
\setlength{\belowdisplayskip}{3pt}
\end{equation} 
where $L_{task}$ is the task-related loss function and 
$\lambda$ is a penalty factor on $L_{memory}$.

\section{MEGA Architecture}

\begin{figure}[t]
  \vspace{-0.2cm}
  \centering
  \includegraphics[scale=0.4,trim=4.7cm 5.1cm 4.9cm 5cm,clip]{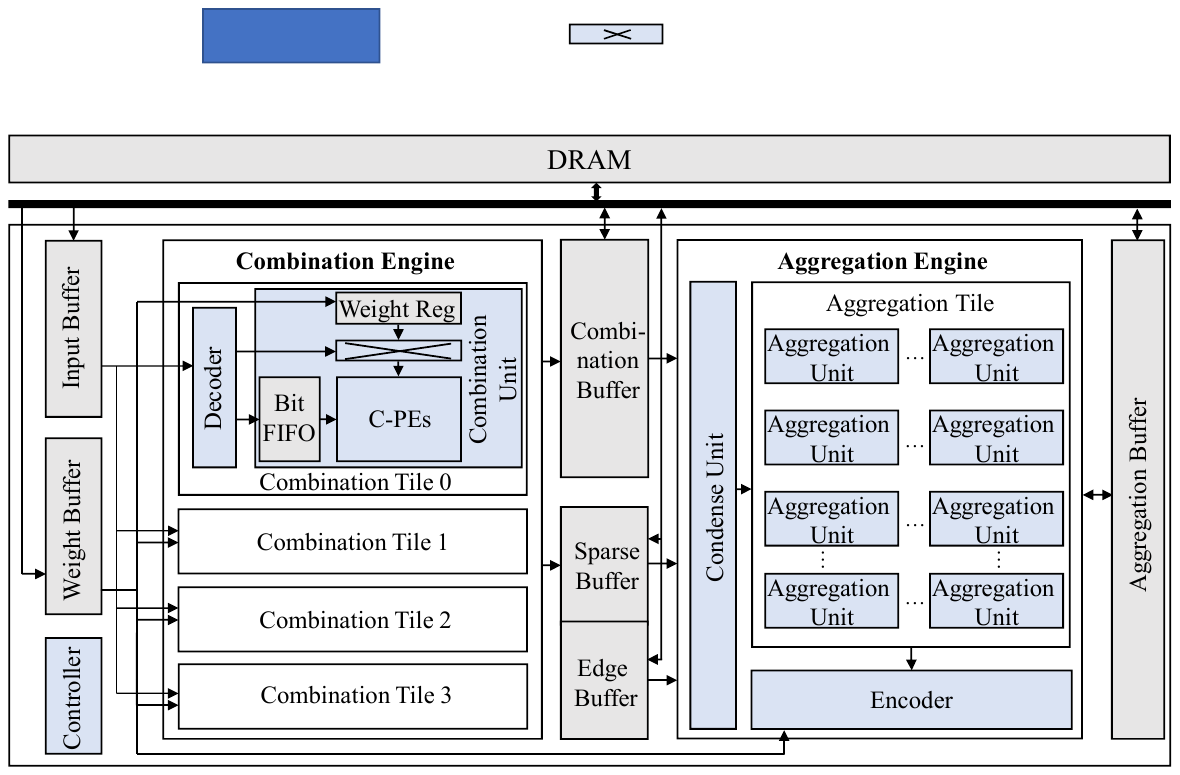}
  \caption{ Overall Architecture of MEGA Accelerator.}
  \vspace{-0.4cm}
  \label{overview}
\end{figure}

\begin{figure*}[ht]
  \vspace{-0.4cm}
  \hspace{-1cm}
  \centering 
  \subfigure[Nodes features]{
    \begin{minipage}[t]{0.25\linewidth}
      \centering
      \includegraphics[scale=0.48,trim=10cm 8.5cm 12.5cm 7.5cm,clip]{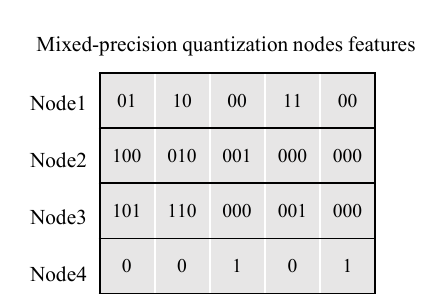}
      \label{node_features}
    \end{minipage}
  }
  \hspace{-0.3cm}
  \subfigure[Bitindex]{
    \begin{minipage}[t]{0.1\linewidth}
      \centering
      \includegraphics[scale=0.48,trim=12.5cm 9cm 12cm 7cm,clip]{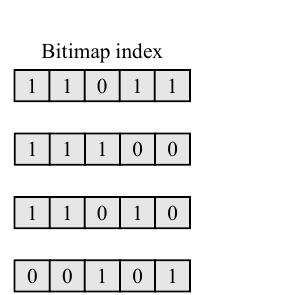}
      \label{bitmap}
    \vspace{-0.4cm}
  \end{minipage}
  }
  \hspace{-0.3cm}
  \subfigure[Fixed length package]{
    \begin{minipage}[t]{0.29\linewidth}
      \centering
      \includegraphics[scale=0.48,trim=10cm 9.7cm 9cm 7cm,clip]{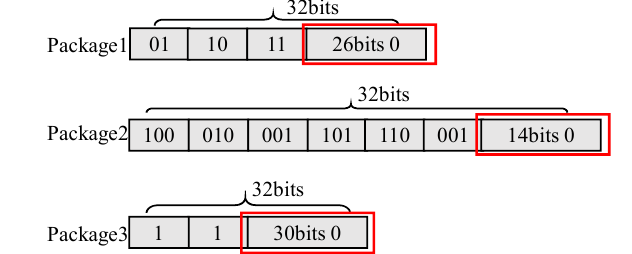}
      \label{adaptive_paddings}
    \vspace{-0.4cm}
  \end{minipage}
  }
  \hspace{-0.3cm}
  \subfigure[Adaptive length package]{
    \begin{minipage}[t]{0.29\linewidth}
      \centering
      \includegraphics[scale=0.48,trim=10cm 9.3cm 7.2cm 7cm,clip]{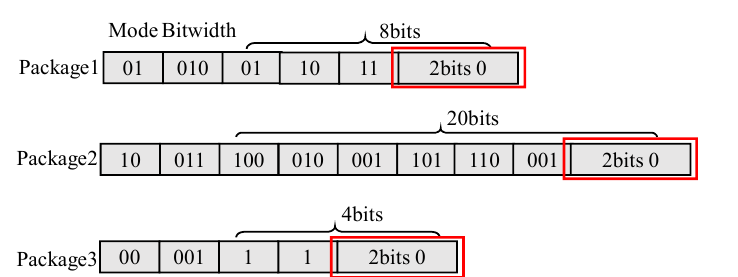}
      \label{adaptive_package}
    \vspace{-0.4cm}
  \end{minipage}
  }
  \vspace{-0.1cm}
  \caption{(a) The fine-grained mixed-precision quantization nodes features. (b) Bitmap index.
    (c) The paddings in the package with fixed length. (d) The Adaptive-Package method in this work.}
  \vspace{-0.4cm}
\end{figure*}

\subsection{Overview}

As illustrated in Fig. \ref{overview}, the accelerator employs a heterogeneous architecture, 
in which the combination
and aggregation phases are performed in the Combination Engine and the Aggregation Engine, 
respectively.
Similar to the previous works\cite{geng2020awb,hwang2023grow,li2021gcnax}, 
MEGA adopts the execution order of $\mathbf{A}(\mathbf{X}\mathbf{W})$ 
to reduce the number of MACs for better efficiency. 
The workflow of MEGA is as follows.

The encoded input nodes and network weights prefetched from the off-chip DRAM are stored 
in the Input and Weight Buffer, respectively. When the calculation starts, the node features 
are read from the Input Buffer and dispatched to the parallel Combination Tiles in the 
Combination Engine, each tile is responsible for processing a slice of features in a node. 
After decoding, the node features in a tile are fed to the Combination Unit and perform the 
combination operation, i.e.,
$\mathbf{B=XW}$, with the corresponding weights.
To exploit the potential benefits introduced by the high sparsity of nodes features, 
the Combination Engine employs the row product dataflow,
which is more suitable for the
wide range of sparsity variation in features, as reported in
\cite{hwang2023grow,yoo2023sgcn,liang2020engn}.
The results from different Combination Tiles
can be 
added together to generate the final results 
or partial sums of one node
once slices in different tiles all have been processed done, 
which enhances the temporal locality.

During aggregation, we divide the graph into several subgraphs and
perform aggregation one subgraph after another.
Once a node finishes the combination phase, the combined node features are sent 
to the Aggregation Engine to perform aggregation.
These features are simultaneously stored in the Combination Buffer or reordered in the 
Sparse Buffer by the Condense 
Unit through
our proposed Condense-Edge scheduling strategy,
which enhances data locality and makes memory access contiguous when 
aggregating other subgraphs.
Unlike the Combination Engine, the Aggregation Engine employs the outer 
product dataflow to reuse the generated node features when calculating $\mathbf{AB}$, where the adjacency matrix $\mathbf{A}$ is stored in the Edge Buffer.
Partial sums of aggregation are 
stored in the Aggregation Buffer.
The aggregated nodes
are quantized and encoded using the Encoder and can be forwarded to the
Combination Engine to fuse the two phases. 
Note that all the buffers use the ping-pong mechanism for pipeline processing.

\subsection{Adaptive-Package Storage Format}
\label{adaptive_package_section}

As illustrated in Section \ref{challenges on accelerator design}, 
the nodes features with mixed precision and 
diverse sparsity introduce significant hardware 
overhead when using existing sparse data representations.
To address this issue, we propose 
an Adaptive-Package format to efficiently 
compress the node features. 

In this method, a package is the primitive unit to store data, which 
consists of three segments: Mode (2bits), Bitwidth (3bits), and
Val Array (adaptive), in which Mode is used to identify 
the length mode (short, medium, long) of Val Array and Bitwidth denotes the 
bitwidth (ranging from 1 to 8bit)
of the values stored in the current package.
To leverage the sparsity inherent in node features,
the Val Array only stores non-zero values of the node features. 
Each node has its own bitmap indices to record the locations of the non-zero values, which 
are stored separately from the packages,
as shown in Fig. \ref{bitmap}.
To reduce the decoding complexity and maximize memory efficiency, we make the following 
design choices:

\textbf{Shared Bitwidth in Each Package.}
We restrict the features in a package to share the same quantization bitwidth, 
identified by the Bitwidth. 
This can effectively reduce the decoding complexity as the arrangement of non-zero values 
is fixed when a package is 
retrieved from memory.
Each package stores the non-zero values of successive 
nodes until the package is full or the quantization bitwidth of the node changes.

\textbf{Adaptive Length of Val Array.}
If the package is not full but the bitwidth of the node changes, 
a new package is needed and 
the remaining bits 
in the current package should be padded with zeros to keep memory alignment.
In this case, it is unsuitable to use the fixed-length Val Array, 
which will introduce severe memory overhead
of paddings, especially for high sparsity or low quantization bitwidth, 
as shown in Fig. \ref{adaptive_paddings}.
Considering the various sparsity and the fine-grained mixed-precision features 
(most are 2/3bit), 
we set the 
package length to be adaptive with three levels, i.e., short, medium, 
and long, which is selected based on the sparsity and quantization 
bitwidth of each node.
The length is indicated by the Mode segment, with \underline{00} for short, 
\underline{01} for medium, and
\underline{10} for long.
For low bitwidth or high sparsity, a lower-level mode is adopted to 
reduce memory overhead.
As the example in Fig. \ref{adaptive_package},
by using the short mode, we can reduce 
the padding from 26bits
to 2bits in Package1, substantially improving memory efficiency.
We empirically set (short, medium, long) = (64bits, 128bits, 192bits) to minimize the storage overhead.

\begin{figure*}[t]
  \vspace{-0.2cm}
  \centering 
  \hspace{-0.5cm}
  \subfigure[The Decoder]{
    \vspace{-0.4cm}
    \begin{minipage}[t]{0.26\linewidth}
      \centering
      \includegraphics[scale=0.4,trim=7cm 4.8cm 10cm 6cm,clip]{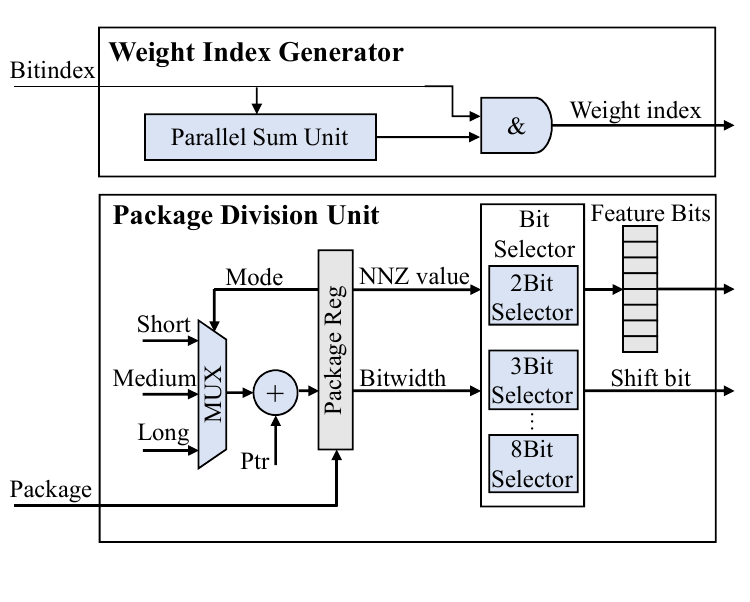}
      \vspace{-1cm}
      \label{decoder}
    \end{minipage}
  }
  \hspace{0.5cm}
  \subfigure[Generate weights indices]{
  \begin{minipage}[t]{0.23\linewidth}
      \centering
      \includegraphics[scale=0.45,trim=9.3cm 6cm 12.3cm 6cm,clip]{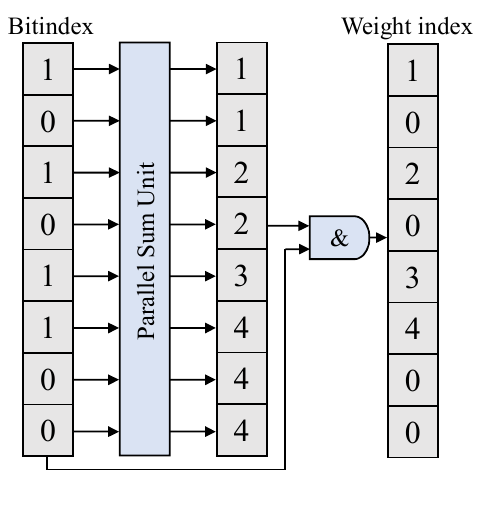}
      \vspace{-0cm}
      \label{weight_index}
    \end{minipage}
  }
  \subfigure[The Combination Unit]{
  \begin{minipage}[t]{0.45\linewidth}
      \centering
      \includegraphics[scale=0.5,trim=7cm 6.7cm 7cm 6.1cm,clip]{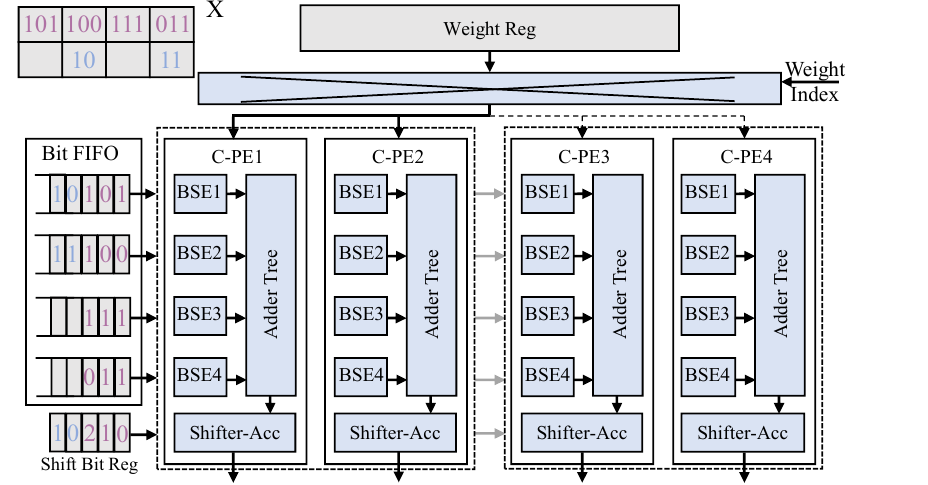}
      \label{combination_unit}
    \end{minipage}
    \hspace{-0.5cm}
  }
  \vspace{-0.1cm}
  \caption{ Microarchitecture of the Combination Engine including a Decoder and a Combination Unit.
  }
  \vspace{-0.3cm}
\end{figure*}

\subsection{Combination Engine}
\textbf{\textit{1) Decoder:}}
As illustrated in Fig. \ref{decoder}, the Decoder is composed of a Package Division Unit 
and a Weight Index Generator, in which
the goal of the Package Division Unit is to select valid values within 
the package, and the Weight Index Generator is
responsible for calculating the row indices used to load weights. 
The Package Reg is a register used to cache the package from Input Buffer.
First, the Package Division Unit extracts 
the Mode and Bitwidth 
from the Package Reg and 
feeds the non-zero values within the package to 
the Bit Selector. 
Next, the Bit Selector selects the valid bits using the corresponding selector, 
and then passes them to the Combination Unit for bit-serial computation.
For example, 
when the Val Array of a package is $10,01,10,11$ and the quantization bitwidth is 2bit, 
the 2Bit Selector is used to select the valid bits as 
$1\underline{0},0\underline{1},1\underline{0},1\underline{1}$, and generates the 
pairs of feature bits and shift bit (0101,0 and 1011,1), which are 
fed to the Combination Unit to be 
processed in sequence.
Shift bit is used to control the number of shifts in bit-serial computation.
Since features in each package use the same bitwidth, 
the data format is fixed, which can substantially 
decrease the hardware complexity of the bit selectors.  
The Weight Index Generator parses the bitindex of node features 
in parallel by the Parallel Sum Unit 
to convert the bitindex to indices of the rows in $\mathbf{W}$, 
as shown in Fig. \ref{weight_index}. 
Then these decoded weight indices are fed to the 
Combination Unit to load weights.

\begin{figure}[t]
  \vspace{-0cm}
  \centering
    \includegraphics[scale=0.38,trim=6cm 8.3cm 2.3cm 4.7cm,clip]{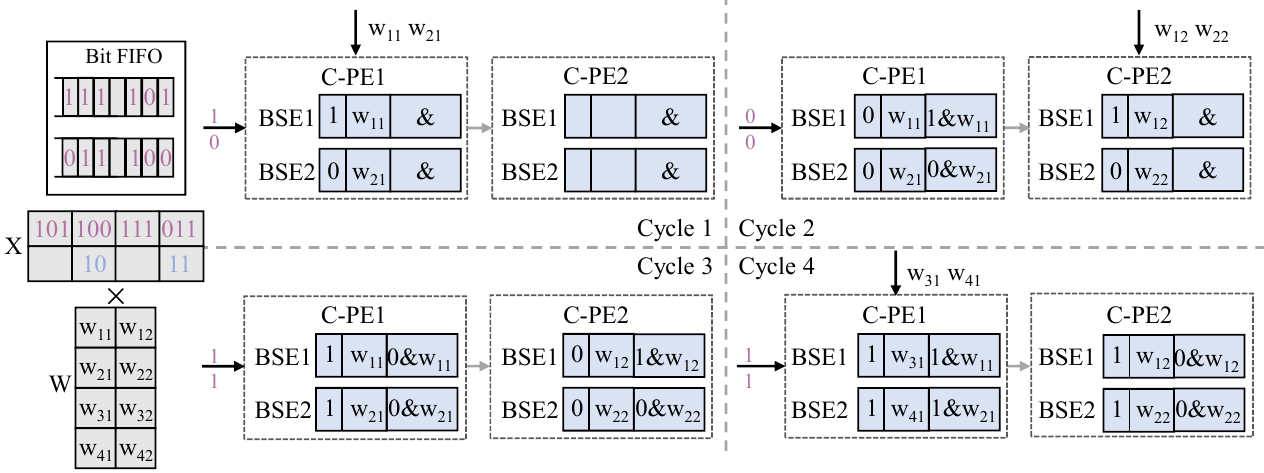}
    \vspace{-0.1cm}
    \caption{The cycle by cycle execution of two C-PEs, in which two 
    BSEs process a group of two bits.}
    \label{combination_pipeline}
    \vspace{-0.5cm}
\end{figure}

\textbf{\textit{2) Combination Unit:}}
The Combination Unit 
adopts the row-product manner and bit-serial computation approach
to complete mixed-quantization calculation 
of $\mathbf{B=XW}$ ($\mathbf{X}$ is from 1 to 8bit).
As shown in Fig. \ref{combination_unit}, a Combination Unit 
composes of a Bit FIFO, a Weight Reg, 
and $\#m$ Combination PEs (C-PEs). 
Node features from the 
Decoder are stored in the Bit FIFO and are subsequently 
loaded into the C-PEs in a bit-serial manner. 
The dense weights (quantized to 4bit) are buffered in the Weight Reg and are fetched 
to C-PEs via a crossbar
according to the row indices from the Decoder.
The C-PEs are responsible for the mixed-precision computation, and 
$\#m$ C-PEs can complete a Vector-Matrix multiplication between one row of $\mathbf{X}$ and 
$m$ columns of $\mathbf{W}$. Different C-PEs calculate different Vector-Vector 
multiplication to 
get $m$ different output features in the same row of $\mathbf{B}$.
Each C-PE comprises of $\#n$ Bit-Serial Engines (BSEs), an 
Adder Tree, 
and a Shifter-Acc. Different BSEs complete the multiplication of 
different non-zero elements in the same row of $\mathbf{X}$
using corresponding weights.
Since the bit-serial method greatly simplifies the 
multiplication operation, BSE only consists of an AND unit and three registers, including
a weight register, a feature bit register, and a result register.

During processing, the C-PEs are equally divided into two groups, as illustrated in 
Fig. \ref{combination_unit}, 
and C-PEs in each group share the same features. 
The feature bits in the Bit FIFO are fed into the C-PEs 
for calculation and flow from the left half C-PE group to the right half for data reuse. 
Within each C-PE, BSEs execute AND operations between the 
feature bits and the loaded weights, with the results subsequently 
summarized in the Adder Tree. Then the sum is shifted according to the 
shift bit from the Decoder and accumulated with the
partial sum until the 
final output is produced. 
In this way, the weights are reused across different bits of the 
node features. Given that the minimum bitwidth of the node features 
is 2bit in Degree-Aware method, each weight is 
reused at least twice, which reduces the weight bandwidth requirement of the C-PEs 
by half
and also decreases the hardware complexity of the crossbar.
Because the density of $\mathbf{W}$ is
always 100\%, the BSEs in the same row are all busy during the systolic process.
A $c\times n(\frac{m}{2}\times 4\ \text{bits})$ crossbar 
is used to unicast corresponding rows of $W$ to C-PEs, where $c$ is 
the dimension of the feature slice processed in each Combination Tile.

Fig. \ref{combination_pipeline} shows an example of the dataflow with m=2 and n=2.
In the first cycle, the first feature bits and the corresponding weights 
are loaded to BSE1 and BSE2 of C-PE1. In the second cycle, BSE1 and BSE2 of C-PE1
perform the AND operations and load the second bits of the same non-zero values.
The crossbar only loads weights to C-PE2, 
because
the weights loaded to C-PE1 in the first cycle can be reused by the different 
bits of the same non-zero values.
Moreover, the bits in C-PE1 are forwarded to C-PE2 for subsequent processing. 
The third 
and fourth cycles perform similarly.

\begin{figure*}[t]
  \vspace{-0.2cm}
  \centering 
  \subfigure[]{
  \hspace{-1.1cm}
  \begin{minipage}[t]{0.12\linewidth}
    \centering
      \includegraphics[scale=0.35,trim=12cm 7.7cm 10.5cm 6.8cm,clip]{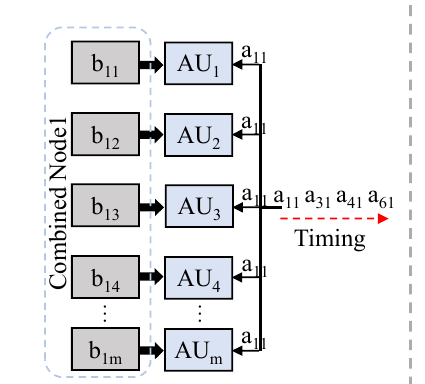}
      \label{aggregation_unit}
    \vspace{-0.8cm}
  \end{minipage}
  }
  \subfigure[]{
    \begin{minipage}[t]{0.23\linewidth}
    \centering
      \includegraphics[scale=0.35,trim=8cm 7.8cm 11cm 6cm,clip]{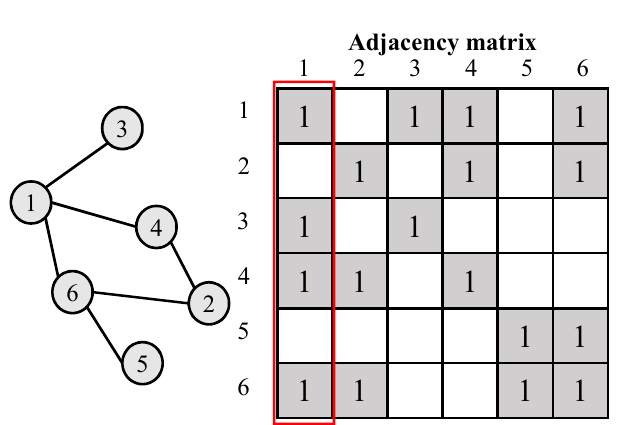}
      \label{raw_graph}
    \vspace{-0.8cm}
  \end{minipage}
  }
  \subfigure[]{
  \hspace{-1.3cm}
  \begin{minipage}[t]{0.42\linewidth}
      \centering
      \includegraphics[scale=0.35,trim=5cm 7.9cm 7.3cm 6cm,clip]{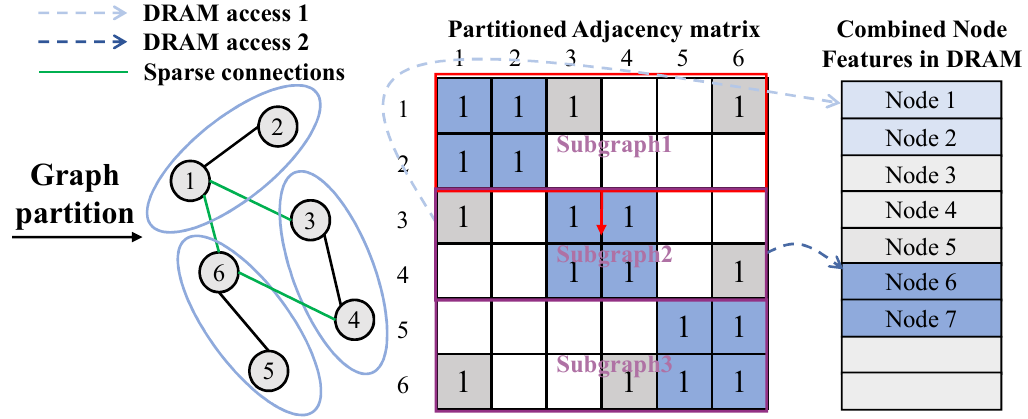}
      \vspace{-1cm}
      \label{partition_graph}
  \end{minipage}
  \vspace{-0.5cm}
  }
  \subfigure[]{
    \hspace{-1.1cm}
    \begin{minipage}[t]{0.24\linewidth}
      \centering
      \includegraphics[scale=0.32,trim=4.9cm 7.4cm 9.3cm 5.8cm,clip]{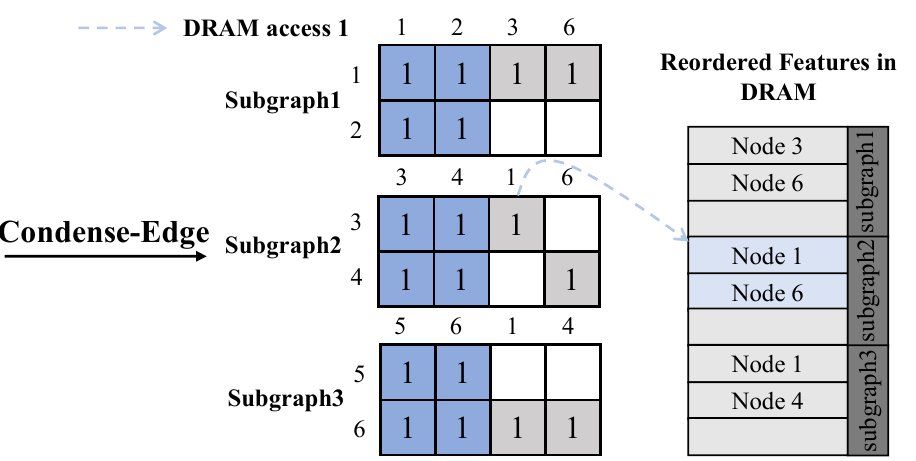}
      \label{condense_graph}
    \vspace{-1.8cm}
  \end{minipage}
  }
  \vspace{-0.1cm}
  \caption{ 
  (a) Aggregate using Node1.
  (b) Baseline graph without partition.
  {(c) DRAM access for aggregating Subgraph2 after partition.}
  (d) DRAM access for aggregating Subgraph2 with the Condense-Edge scheduling strategy.}
  \vspace{-0.2cm}
\end{figure*}

\subsection{Aggregation Engine}

The Aggregation Engine is composed of an Encoder, a Condense Unit, and an Aggregation Tile.
The Condense Unit is used to perform our Condense-Edge scheduling strategy (detailed in Section
\ref{condense_edge_section}).
The Aggregation Tile contains multiple 
Aggregation Units (AUs), which can perform integer addition and multiplication of scalar.
Considering the results (quantized to 4bit) 
from Combination Engine are node-by-node and nearly 100\% dense,
the Aggregation Engine calculates the $\mathbf{A}\mathbf{B}$ in an outer product manner
to fully reuse the generated nodes features. 
Different dimensions of features in a node are unicasted to different AUs and multiplied with 
the corresponding
edge weight of $\mathbf{A}$ broadcasted to each AU, which is stored in the Edge Buffer using 
the CSC format.
As the example shown in Fig. \ref{aggregation_unit} and \ref{raw_graph}, the combined
Node1 is used to partially aggregate Node1, Node3, Node4, and Node6 in sequence.
If the features of one node 
cannot occupy all units, free units can be assigned to aggregate other nodes. 
The multiplied results are added 
with 16bit partial sums buffered in Aggregation Buffer to 
complete the aggregation.
When the aggregation for a node is complete, it is sent to the Encoder to be encoded to 
Adaptive-Package format.

The Encoder is composed of $\#m$ Quantization-Nonlinear (QN) units,
which
complete the quantization process of $\#m$ values 
using the scales from the Weight Buffer.
Moreover, each QN unit performs the ReLU non-linear 
operation
and generates a bitindex to 
identify whether the value is zero. 
The non-zero values are concatenated and stored in a package 
register as the Val Array in a package. 
To maximize memory efficiency,
{we adopt a greedy heuristic method where
the package register continuously stores non-zero values 
in successive nodes until the maximum package length is reached or 
the node's bitwidth
changes.} Then we select the appropriate Mode according
to the bit length of the 
values in the package register. 
And the Mode and Bitwidth
are assembled as the header information and the package is written back into 
the Input Buffer as the input of the combination phase of the next layer. 
{The Encoder repeatedly performs the above process until the aggregation phase of the current layer 
completes.}

\setlength{\textfloatsep}{10pt}
\begin{algorithm}[t]

  \scriptsize
	\renewcommand{\algorithmicrequire}{\textbf{Initialize:}}
	\renewcommand{\algorithmicensure}{\textbf{Output:}}
	\caption{Condense-Edge scheduling strategy}
	\label{alg1}
	\begin{algorithmic}[1]
      \STATE \textbf{Initialize:} $\mathbf{eID\_list=\{eID\_FIFO_1,...,eID\_FIFO_n\}}$: 
      container 
      of the sparse connections $e\_IDs$ and $eID\_FIFO_i$ stores $e\_IDs$ of the $i$-th subgraph 
      in ascending order;
      $\mathbf{Address\_list}$: stores the initial memory
      addresses of the Sparse Buffer corresponding to different subgraphs.
      \STATE \textbf{Outputs:} None (reorder the nodes features)
      \STATE \textbf{Condense-Edge}($x$, $nID$, $SubNum$):
      \STATE Partition the graph using METIS and initialize $\mathbf{eID\_list}$. $subID=0$
        \STATE Store $x$ into the Combination Buffer
        \WHILE{$subID< SubNum$}
          \STATE $eID\_FIFO = eID\_list[subID]$ 
          \IF {$eID\_FIFO[0] == nID$}
            \STATE  $eID\_FIFO.pop()$\COMMENT{\textcolor{blue!50}{\# invalid the $eID$ have matched}}
            \STATE $ptr = Address\_list[subID]$ and store the node features $x$ 
            into Sparse Buffer according to the $ptr$
            \COMMENT{\textcolor{blue!50}{\# reorder the nodes features in Sparse Buffer}}
            \STATE Increment $ptr$ and $Address\_list[subID]=ptr$
            \STATE $subID = subID + 1$ \COMMENT{\textcolor{blue!50}{\# compare with the next subgraph}}
          \ELSE
            \STATE $subID = subID + 1$
          \ENDIF
          \STATE If the corresponding region of the Sparse Buffer is full, 
          write the data back to DRAM and reinitialize 
          the $ptr$.
        \ENDWHILE
      \STATE \textbf{end}
   \end{algorithmic}
\end{algorithm}
\setlength{\textfloatsep}{10pt}

\subsection{Condense-Edge Scheduling Strategy}
\label{condense_edge_section}

During aggregation, the extreme sparsity of $\mathbf{A}$ ruins the data locality and 
its large size
results in a considerable amount of 
partial sums, which puts forward high demand on the capacity of the Aggregation Buffer.
Therefore, to improve the data reuse and reduce the resource requirement,
many modern GNN accelerators 
divide the graph into several subgraphs using 
METIS\cite{abou2006multilevel}, and then perform the aggregation one subgraph after another
\cite{hwang2023grow,you2022gcod}, 
as shown in Fig. \ref{partition_graph}. 
In this way, the Aggregation Buffer only stores the aggregated partial sums of the 
processing subgraph and the spatial locality in the subgraph {(blue region 
in adjacency matrix)} is 
significantly improved.

However, as our analysis in Section \ref{challenges on accelerator design},
the pitfall for the partition is the more irregular sparse connections 
{(grey region in the adjacency matrix of Fig. \ref{partition_graph})} 
between subgraphs,
which engenders the bottleneck in advancing performance.
As an example shown in Fig.\ref{partition_graph},
when using the sparse connections to aggregate the nodes in Subgraph2, 
DRAM accesses 
are needed to read the features of Node1 {(to aggregate Node3)} 
and Node6 {(to aggregate Node4)}. 
{Assuming that the granularity of each
DRAM access is 128B, i.e., a continuous memory region of two nodes where 
each node has 128-dimension features and is quantized to 4bit, }
then the access behavior is:
Node1 \textbf{miss}$\rightarrow$Node1\&2 loaded from 
DRAM$\rightarrow$Node6 \textbf{miss} 
$\rightarrow$Node6\&7 loaded from DRAM,
which requires two DRAM accesses 
and only half is utilized in each DRAM access.
To alleviate the irregularity induced by sparse connections,
we propose the Condense-Edge scheduling strategy.

{MEGA also performs the aggregation phase
one subgraph after another. The key idea of the Condense-Edge scheduling strategy is that 
when the Aggregation Engine performs aggregation of the first 
subgraph using the newly combined node, the Condense Unit simultaneously reorders the node features.
Then, the features required by the sparse connections is stored continuously, which will be used to aggregate other subgraphs, 
significantly improving DRAM access utilization.}

As shown in Algorithm \ref{alg1}, we first divide the graph into several subgraphs 
using METIS\cite{abou2006multilevel}.
Since partition is performed offline, we can obtain the number of subgraphs $SubNum$ and 
sparse connection IDs ($eIDs$) of each subgraph in advance, e.g., 3 and 6 in Subgraph1, 
which 
are stored in the Edge Buffer.
We initialize the $eID\_list$ using these IDs.
{The Sparse Buffer is divided into $SubNum$ regions and 
each region stores the 
nodes features used by sparse connections of one specific subgraph. The $\mathbf{Address\_list}$ is initialized by the 
initial addresses of these regions.}
Once the features $x$ of one node generate from the Combination Engine, 
we compare its ID $nID$ with $eIDs$ of different subgraphs. 
If there is a match, it implies that the node features will be utilized by this sparse connection,
and thus we store $x$ into the region of Sparse Buffer 
assigned 
to the corresponding subgraph ($subID$) 
according to the address stored in $Address\_list$.
We then update the address to ensure the continuity of nodes features in the Sparse Buffer
(line 11).
{When multiple nodes within the same subgraph are all connected with a node $i$ 
through sparse connections, 
we only store node $i$ once because it can be reused by these nodes 
in the same subgraph.}

To curtail the overhead of matching, we store 
$eIDs$ in ascending order and initially set all IDs to be valid. 
If the ID matches, 
we set this ID to be invalid (line 9). 
In this way, once the node is generated from the Combination Engine,
we only need to compare with the first ID in each $eID\_FIFO$ 
instead of all IDs, which considerably reduces 
the matching overhead.  
Regardless of whether there is a match or not, 
the node features need to be written to the Combination Buffer to 
aggregate the nodes without sparse connections.

\begin{figure}[t]
  \vspace{-0cm}
  \centering
  \includegraphics[scale=0.53,trim=7cm 6.5cm 6.2cm 6.9cm,clip]{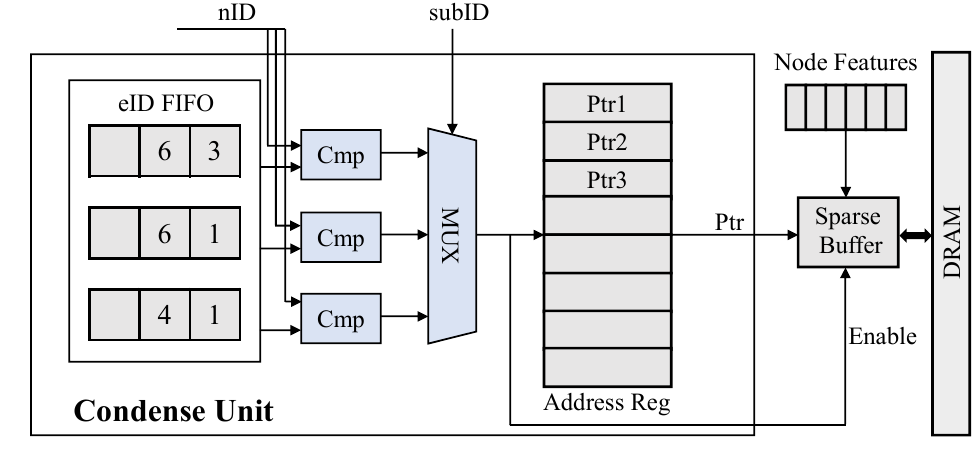}
  \vspace{-0cm}
  \caption{Microarchitecture of the Condense Unit.}
  \vspace{-0cm}
  \label{condense_unit}
\end{figure}
When aggregating other subgraphs, 
the Sparse Buffer 
reads the nodes features used by sparse connections
that have been continuously stored in DRAM.
{When features of one node is ready to be fetched to aggregate the nodes in the current subgraph, 
its $nID$ is compared with the $eIDs$. }
If there is a match, the Condense Unit directly fetches the required features from 
the Sparse Buffer. Otherwise, the needed node features are read from 
the Combination Buffer. As shown in Fig. \ref{condense_graph}, 
the access behavior of using sparse connections in Subgraph2 turns to: {Node1 \textbf{miss}$\rightarrow$Node1 is used by the 
sparse connection$\rightarrow$Node1\&6 loaded from DRAM to the Sparse Buffer$\rightarrow$Node6 match$\rightarrow$fetch Ndoe6 
from the Sparse Buffer.} 
Our Condense-Edge 
strategy reduces DRAM access 
from 2 to 1, improving the 
data reuse of each DRAM access.

Fig. \ref{condense_unit} displays the microarchitecture of the Condense Unit.
Each eID FIFO has eight entries and stores the $eIDs$
of a single subgraph, which are arranged in ascending order.
{Moreover,
each eID FIFO is equipped with a comparator} (Cmp) to improve the parallelism of matching.

\section{Evaluation}

\subsection{Experiments Setup}
\label{Experiments Setup}
\textit{\textbf{1) Models and Datasets:}}
We evaluate our framework on three GNN models - 
GCN\cite{kipf2016semi}, GIN\cite{xu2018powerful}, and
GraphSage\cite{hamilton2017inductive} 
using five datasets 
including three citation datasets (Cora, Cite, PubMed)\cite{yang2016revisiting}, a knowledge 
graph dataset (NELL)\cite{carlson2010toward}, and a large scale dataset 
Reddit\cite{hamilton2017inductive}.  
Details about datasets and models are 
shown in TABLE \ref{sta_dataset} and TABLE \ref{model}, respectively. 

\textit{\textbf{2) Baselines:}}
At the algorithm level, 
we compare our Degree-Aware
quantization method with the FP32 models and the prior art method DQ\cite{tailor2020degree} 
with 4bit (DQ-INT4) on various tasks.
As node features dominate the memory consumption in GNNs, 
we count the average bitwidths for nodes features of the overall model and take the ratio of the 
average bitwidth / 32 as the theoretical compression ratio, named `CR'.
At the hardware level,
to evaluate our MEGA accelerator,
we compare with four state-of-the-art architectures, i.e., HyGCN\cite{yan2020hygcn}, 
GCNAX\cite{li2021gcnax}, 
GROW\cite{hwang2023grow}, and SGCN\cite{yoo2023sgcn}. 

\setlength{\textfloatsep}{5pt}
\begin{table}[t]\scriptsize
\vspace{-0cm}
  \caption{The statistics of datasets used in this work.}
\vspace{-0.2cm}
\label{sta_dataset}
  \begin{center}
    \begin{tabular}{ccccc}
      \toprule[1.5pt]
      Dataset  & \#Node & \#Edge      & Feature   length & Average   degree \\ \midrule[1pt]
      Cora     & 2,708   & 10,556      & 1,433            & 3.90             \\
      CiteSeer & 3,327   & 9,104        & 3,703            & 2.74             \\
      PubMed   & 19,717  & 88,648      & 500              & 4.50             \\
      NELL     & 65,755  & 251,550     & 61,278           & 3.83             \\
      Reddit   & 232,965 & 114,615,892 & 602              & 491.99           \\ \bottomrule[1.5pt]
    \end{tabular}
  \end{center}
  \vspace{-0.2cm}
\end{table}

\begin{table}[t]\scriptsize
  \vspace{-0.3cm}
  \caption[short]{The GNN models specifications.}
  \vspace{-0.2cm}
  \label{model}
  \begin{center}
    \begin{tabular}{ccccc}
      \toprule[1.5pt]
      Model     & Sample   nodes & Layers & Hidden   units & Aggregation \\ \midrule[1pt]
      GCN       & —              & 2      & 128            & Add         \\
      GIN       & —              & 2      & 128            & Add         \\
      GraphSage & 25             & 2      & 256            & Mean        \\ \bottomrule[1.5pt]
      \end{tabular}
  \end{center}
  \vspace{-0.2cm}
\end{table}

\textit{\textbf{3) Simulation:}}
We use Verilog to implement MEGA
and synthesize the 
RTL with Synopsys Design
Compiler using the TSMC 28nm standard library at 1GHz to obtain area and 
power consumption of the processing units. For the SRAM-based on-chip buffers, 
we use CACTI-7.0\cite{balasubramonian2017cacti} to model the
area and power consumption. In TABLE \ref{chip_config}, we give the configuration
of MEGA and its 
breakdown analysis of area and power. 
We use HBM1.0\cite{o2014highlights} with 256GB/s 
bandwidth as the DRAM and model the off-chip access using 
Ramulator\cite{kim2015ramulator} and estimate the energy 
consumption for accessing DRAM following the methodology in \cite{yan2020hygcn}.

\linespread{1}
\setlength\tabcolsep{3pt}
\begin{table}[t]\scriptsize
  \vspace{-0cm}
  \caption{The configuration and breakdown of MEGA.}
  \vspace{-0.2cm}
  \label{chip_config}
  \vspace{-0cm}
  \begin{center}
    \begin{tabular}{ccccc}
      \toprule[1.5pt]
                                        & Component            & Area($mm^2$)& Power($mW$)  & Config       \\ \midrule[1pt]
      \multirow{7}{*}{\begin{tabular}[c]{@{}c@{}}Processing \\ Unit\end{tabular}} & BSEs  & 0.053       & 14.70         & $4\times 8\times 32$       \\
                                        & Aggregation   Unit  & 0.100       & 28.92        & 256          \\
                                        & Crossbar             & 0.027       & 5.56         & {\begin{tabular}[c]{@{}c@{}}$32\times 8$(64bit)\end{tabular}} \\
                                        & Condense   Unit   & 0.002       & 1.19         & 16 ID FIFOs            \\
                                        & Encoder      & 0.010       & 1.81         & 32 QN units            \\
                                        & Decoder   & 0.003       & 0.75         & —            \\
                                        & Others               & 0.004       & 0.80         & —            \\ \hline
      \multicolumn{2}{c}{Total}                                   & 0.199(11\%) & 53.73(28\%)  & Size (KB)     \\ \hline
      \multirow{6}{*}{\begin{tabular}[c]{@{}c@{}}Data \\ Buffer\end{tabular}}    & Aggregation   Buffer & 0.540       & 46.56        & 128          \\
                                        & Combination   Buffer & 0.452       & 35.19        & 96           \\
                                        & Input   Buffer       & 0.220       & 22.88        & 64           \\
                                        & Edge   Buffer           & 0.119       & 9.44         & 24            \\
                                        & Sparse   Buffer      & 0.154       & 12.86        & 32           \\
                                        & Weight   Buffer      & 0.190       & 14.32        & 48           \\ \hline
      \multicolumn{2}{c}{Total}                                    & 1.67(89\%)  & 141.25(72\%) & 392            \\ \hline
      \multicolumn{2}{c}{Measured Total(28nm)}                & 1.869       & 194.98       & —            \\ \bottomrule[1.5pt]
    \end{tabular}
  \end{center}
  \vspace{-0.2cm}
\end{table}

\linespread{1.1}
\setlength\tabcolsep{3pt}
\begin{table}[t]\scriptsize
  \vspace{-0.2cm}
  \caption{The configurations of compared architectures.}
  \vspace{-0.2cm}
  \label{archi_sta}
  \begin{center}
    \begin{tabular}{cccccc}
      \toprule[1.5pt]
      Accelerator &  \begin{tabular}[c]{@{}c@{}}Computing \\ Unit@1GHz\end{tabular}    &  Area($mm^2$) & Sparsity      & Precision & \begin{tabular}[c]{@{}c@{}}Graph \\ Partition\end{tabular} \\ \midrule[1pt]
      $\text{HyGCN}^{*}$        & \begin{tabular}[c]{@{}c@{}}16 MACs\\ 4 SIMD16\end{tabular} & 1.86  & NO            & 32bits    & No                                                         \\ \hline
      GCNAX        & 32 MACs                                                              & 1.85  & Both   Phases & 32bits    & No                                                         \\ \hline
      $\text{SGCN}^{*}$         & \begin{tabular}[c]{@{}c@{}}16 MACs\\4 SIMD16\end{tabular} & 2.39  & \begin{tabular}[c]{@{}c@{}}Aggregation \\ Phase\end{tabular}   & 32bits    & No                                                         \\ \hline
      GROW         & 32   MACs                                                               & 2.36   & Both   Phases   & 32bits    & Yes                                                         \\ \hline
      MEGA          & \begin{tabular}[c]{@{}c@{}}4x8x32 BSEs\\256 Aggre Units \end{tabular} & 1.87     & Both   Phases   & Mixed     & \begin{tabular}[c]{@{}c@{}}Condense-\\Edge \end{tabular}                                                        \\ \bottomrule[1.5pt]
      \multicolumn{6}{l}{\scriptsize * 16 MACs for combination and 4 SIMD16 for aggregation.} \\
    \end{tabular}
  \end{center}
  \vspace{-0.2cm}
  \end{table}

We implement cycle-accurate simulators of MEGA and the compared 
accelerators
to 
simulate the behavior of the microarchitecture with the same DRAM bandwidth 
to obtain the number 
of cycles executed. 
{The execution cycles of the computation modules of all simulators, including 
MEGA and all compared accelerators, 
have been validated with
the HDL design at the cycle level.}
Given the significant configuration differences among compared accelerators, 
e.g., 22MB buffer size in HyGCN while 512KB
in SGCN and a $32\times 128$ array in HyGCN while 16 MACs in GCNAX, 
we make the following efforts to ensure fair comparisons.
First, we use the same on-chip buffer/cache capabilities
(392KB) as MEGA. 
Second, for HyGCN and SGCN, which also employ a heterogeneous architecture, 
we match the 
OPS and 
convert the arithmetic operation to BitOP
for consistency, with a 32bit fixed-point multiplication equivalent to 1024 BitOPs.
Finally, for GCNAX and GROW adopting uniform computing units, we ensure 
the areas of their computing units to be comparable with the total area 
of the Processing Unit in our MEGA during simulation.
More detailed configurations are listed in TABLE \ref{archi_sta}.

\begin{figure*}[t]
\vspace{-0.2cm}
\centering
  \includegraphics[scale=0.6,trim=0.2cm 5.7cm 0.0cm 7cm,clip]{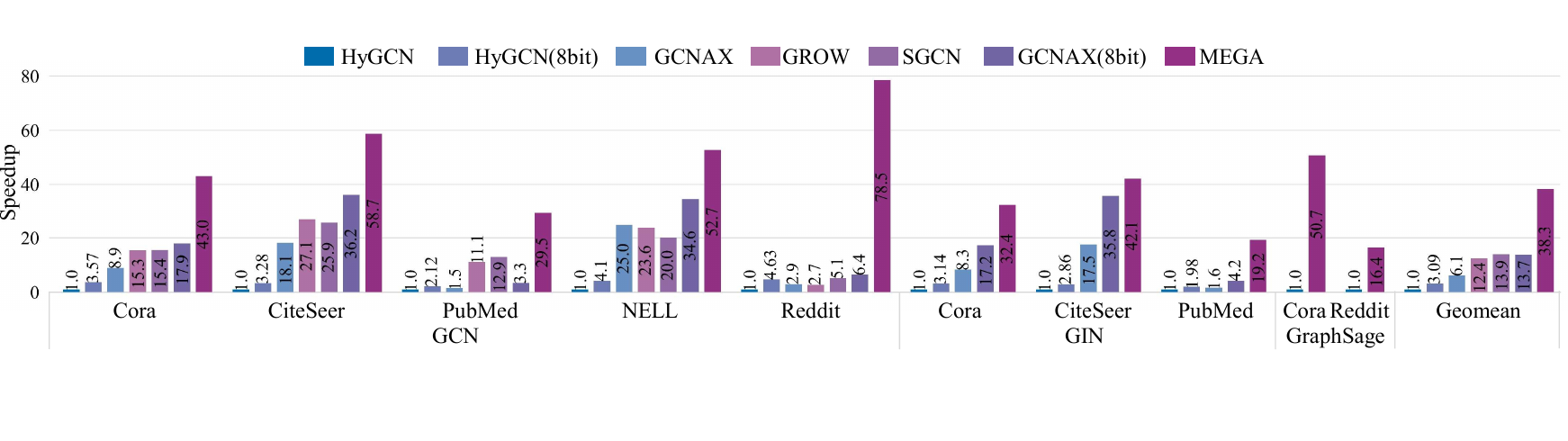}
  \vspace{-1.2cm}
  \caption{Performance comparison with state-of-the-art GNN accelerators (normalized to HyGCN).}
  \vspace{-0.4cm}
  \label{speedup_quantize}
\end{figure*}

\linespread{1}
\setlength\tabcolsep{8pt}
\begin{table}[t]\scriptsize
  \vspace{-0cm}
  \caption{Comparisons with FP32 and DQ-INT4 models.}
  \vspace{-0.2cm}
  \label{quantization_accuracy}
  \begin{center}
    \begin{tabular}{ccccc}
      \toprule[1.5pt]
              Dataset           & Config       & Acc             & Average Bits & CR \\ \midrule[1pt]
      \multirow{6}{*}{Cora}     & GCN(FP32)    & 81.5±0.7\%           & 32            & 1x                                                                \\
                                & GCN(DQ) & 78.3±1.7\%           & 4             & 8x                                                                \\
                                & GCN(ours)    & \textbf{80.9±0.6\%} & \textbf{1.70} & \textbf{18.6x}                                                    \\ \cline{2-5} 
                                & GIN(FP32)    & 77.6±1.1\%           & 32            & 1x                                                                \\
                                & GIN(DQ) & 69.9±3.4\%           & 4             & 8x                                                                \\
                                & GIN(ours)    & \textbf{77.8±1.6\%}  & \textbf{2.37} & \textbf{13.4x}                                                    \\ \cline{2-5} 
                                & GraphSage(FP32)   & 79.7±0.5\%           & 32            & 1x                                                                \\
                                & GraphSage(Ours)   & 79.7±1.0\%           & \textbf{3.40}  & \textbf{9.4x}                                                     \\ \hline
      \multirow{6}{*}{CiteSeer} & GCN(FP32)    & 71.1±0.7\%          & 32            & 1x                                                                \\
                                & GCN(DQ) & 66.9±2.4\%           & 4             & 8x                                                                \\
                                & GCN(Ours)    & \textbf{70.6±1.1\%} & \textbf{1.87} & \textbf{17.0x}                                                    \\ \cline{2-5} 
                                & GIN(FP32)    & 66.1±0.9\%           & 32            & 1x                                                                \\
                                & GIN(DQ) & 60.8±2.1\%           & 4             & 8x                                                                \\
                                & GIN(ours)    & \textbf{65.1±1.7\%}  & \textbf{2.54} & \textbf{12.6x}                                                    \\ \hline
      \multirow{3}{*}{PubMed}   & GCN(FP32)   & 78.9±0.7\%           & 32            & 1x                                                                \\
                                & GCN(DQ)   & 62.5±2.4\%           & 4            & 8x                                                                \\
                                & GCN(Ours)   & \textbf{78.3±0.6\%}  & \textbf{2.50} & \textbf{12.8x}                                                    \\ \hline
      \multirow{2}{*}{Reddit}   & GraphSage(FP32)   & 95.2±0.1\%           & 32            & 1x                                                                \\
                                & GraphSage(Ours)   & \textbf{95.6±0.2\%}  & \textbf{2.74} & \textbf{11.7x}                                                    \\ \bottomrule[1.5pt]
    \end{tabular}
  \end{center}
  \vspace{-0.2cm}
\end{table}

\begin{figure}[t]
  \centering
  \vspace{-0.1cm}
  \includegraphics[scale=0.43,trim=0cm 0cm 0cm 0, clip]{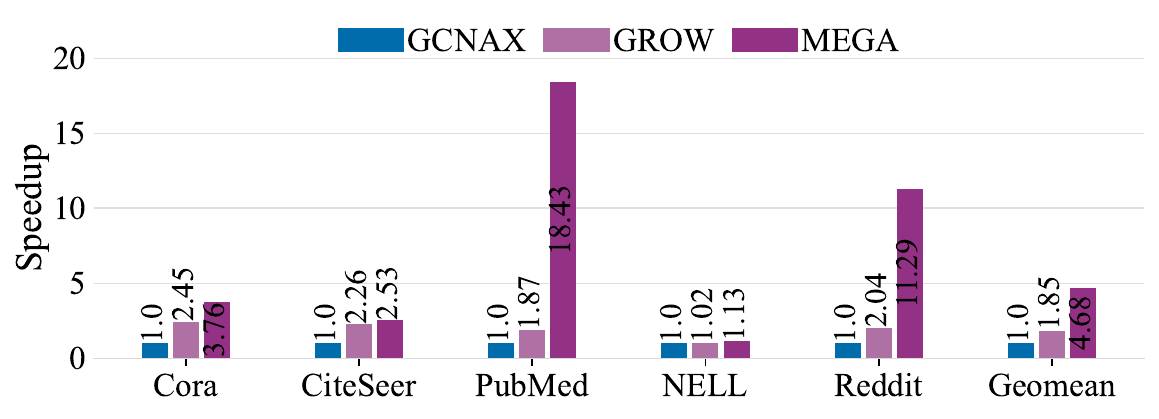}
  \vspace{-0.2cm}
  \caption[short]{{The performance comparison between accelerators 
  using their original configurations and MEGA on GCN (normalized to GCNAX).}}
  \vspace{-0cm}
  \label{ppa}
\end{figure}

\linespread{1.1}
\setlength\tabcolsep{3pt}
\begin{table}[t]\scriptsize
  \vspace{-0.2cm}
  \caption{{The original configurations of compared architectures in a 28nm
  technology node.}}
  \vspace{-0.2cm}
  \label{original_conf}
  \begin{center}
    \begin{tabular}{ccccc}
      \toprule[1.5pt]
      Accelerator &  \begin{tabular}[c]{@{}c@{}}Computing \\ Unit@1GHz\end{tabular}    &\begin{tabular}[c]{@{}c@{}}On-chip \\ Buffer(KB)\end{tabular} &  Area($mm^2$) & Power($mW$)      \\ \midrule[1pt]
      GCNAX        & 16 MACs                                                              &580 & 2.34  & 223.18 \\ \hline
      GROW         & 16   MACs                                                               &538 & 2.67   & 242.44  \\ \bottomrule[1.5pt]
    \end{tabular}
  \end{center}
  \vspace{-0.2cm}
  \end{table}

\subsection{Accuracy}
\label{Evaluation of Degree-Aware Algorithm}
In TABLE \ref{quantization_accuracy}, we present comparisons of the accuracy and 
compression ratio of our method with both FP32
models and DQ-INT4 models (denoted by `DQ') on various tasks. 
Extensive experiments demonstrate that our approach can obtain a high compression ratio while
maintaining accuracy.
Compared with DQ-INT4, our method achieves significantly better accuracy on all tasks, 
even with a higher compression ratio.
Furthermore, we obtain a compression ratio of up to $18.6\times$ over the FP32 
model with negligible loss of accuracy.

We do not compare with DQ-INT8 because our Degree-Aware method is 
able to obtain comparable accuracy compared to DQ-INT8 with 
much higher compression ratios, e.g., DQ-INT8 achieves 71.0\%
with 8bit while Degree-Aware is 70.6\% with 1.87bit on CiteSeer dataset of GCN model.
Moreover, we can trade off accuracy with compression ratio by tuning down the penalty factor 
on the $L_{memory}$. For example, when we turn down the penalty factor, we can achieve 
71.2\% accuracy with 2.24bit, which is still much lower than 8bit.

\subsection{Overall Results}
\label{Overall Results}
\textit{\textbf{1) Speedup:}} 
As shown in Fig. \ref{speedup_quantize}, MEGA on-average achieves 
$38.3\times$, $7.1\times$, $4.0\times$, $3.6\times$ speedup
over HyGCN, 
GCNAX, GROW, and SGCN, respectively.
By supporting the calculation and storage of mixed-precision features, MEGA 
significantly reduces the computation latency and DRAM communication, 
achieving substantial speedups compared to state-of-the-art accelerators.
The speedup over HyGCN is particularly significant because HyGCN 
adopts the execution order of $\mathbf{(AX)W}$, which introduces a large number of 
additional MAC operations, and it does not utilize any sparsity in the 
node features, resulting in considerable redundant DRAM accesses.
Compared to GCNAX, 
MEGA achieves a significant speedup thanks to the lower quantization bitwidth
and fewer pipeline stalls induced by off-chip traffic.
Although GROW and SGCN leverage the sparsity, MEGA 
also obtains significant speedup from the
high compression ratio
brought by our Degree-Aware method. 
{To demonstrate the superiority of our MEGA, we also
compare with the 8-bit version of 
HyGCN and GCNAX, which
run 
GNNs quantized by DQ-INT8 using 8bit MACs, referred to as `HyGCN(8bit)' and `GCNAX(8bit)'. 
We can observe that the improvement over their own 32-bit versions is marginal
because the reduction on DRAM access of quantizing to 8bit is not sufficient for such a 
large number of irregular DRAM accesses in GNN workloads.
On average, the dedicated MEGA significantly outperforms
GCNAX(8bit) on performance by $2.8\times$, which identifies that naively 
replacing the computation units and running 8-bit quantized models on prior accelerators 
are sub-optimal.}

As shown in Fig. \ref{ppa}, to present a more comprehensive comparison, 
we also compare MEGA with GCNAX and GROW that use their original 
configurations reported in corresponding papers. {TABLE \ref{original_conf} presents 
the original configurations of
these compared accelerators.
We do not provide the comparison
with HyGCN and SGCN, which adopt much larger on-chip memory (e.g., 22MB v.s. 392KB) 
or many more computing units (e.g., 32$\times$128 32-bit MAC array v.s. 
4$\times$8$\times$32 BSEs).
With much smaller chip area and power, MEGA on-average obtains 
$4.68\times$ and $2.53\times$ performance over GCNAX and GROW, respectively, thanks to the
significant reduction in DRAM accesses and computational latency 
brought by our software and hardware co-design framework.}

\begin{figure}[t]
  \vspace{-0.1cm}
  \centering 
  \begin{minipage}[t]{1\linewidth}
    \centering
    \includegraphics[scale=0.33,trim=2cm 6.5cm 2cm 6.8cm,clip]{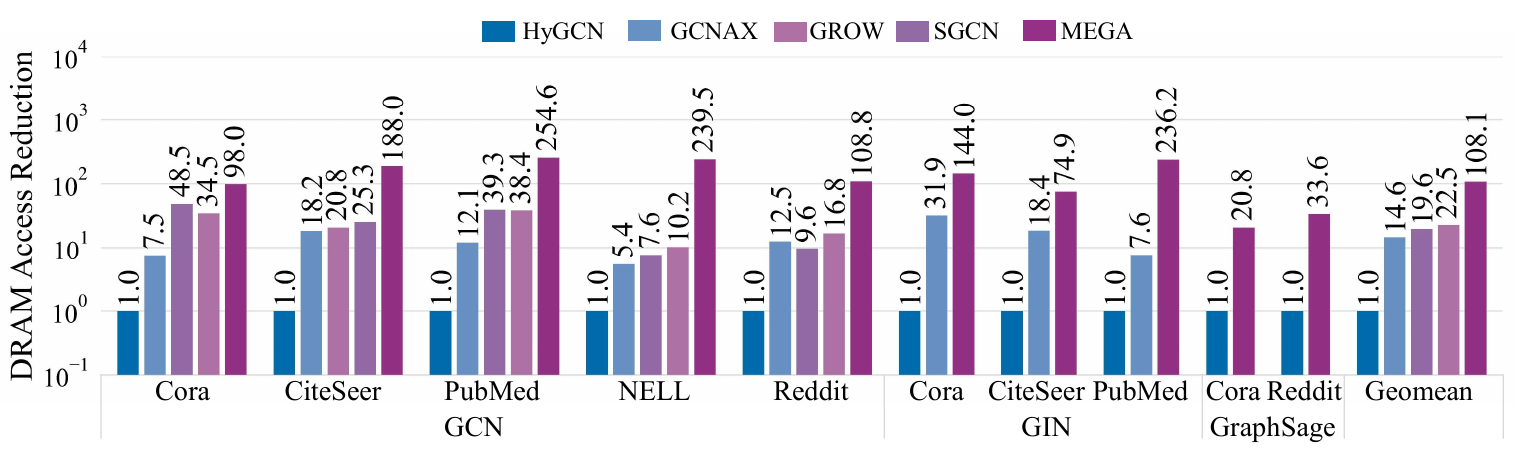}
    \vspace{-0.2cm}
    \caption[short]{{DRAM access reduction of MEGA and the baseline accelerators 
    (normalized to HyGCN).}}
    \vspace{-0cm}
    \label{dram_access}
  \end{minipage} 
  \\
    \begin{minipage}[t]{1\linewidth}
    \centering
    \includegraphics[scale=0.33,trim=2cm 6.5cm 2cm 6.8cm,clip]{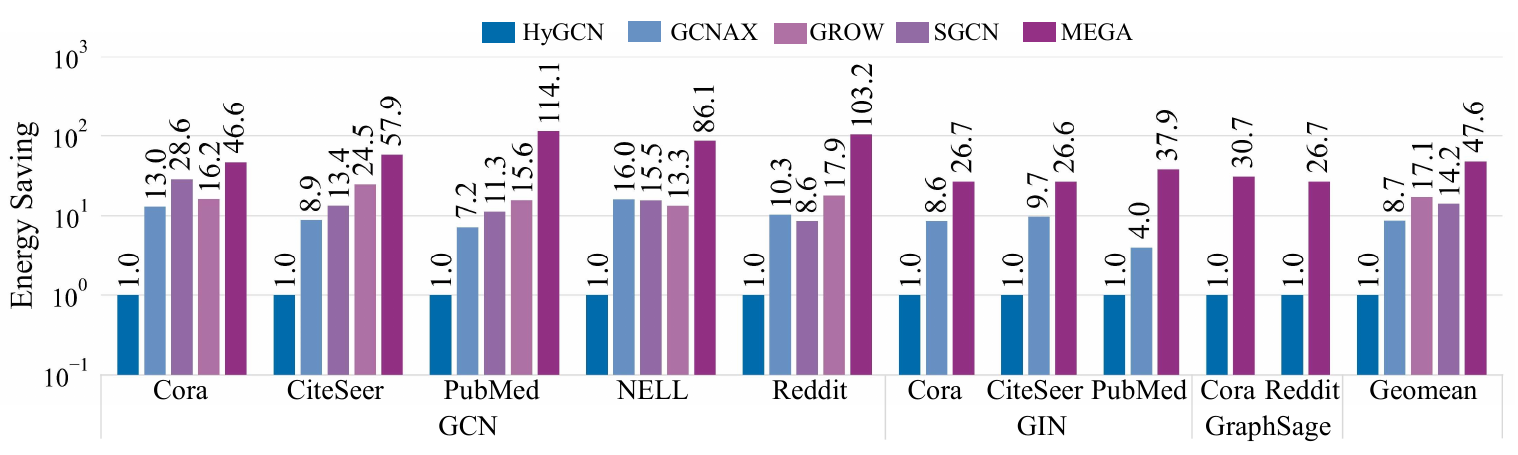}
    \vspace{-0.2cm}
    \caption[short]{{Energy savings of MEGA over the baseline accelerators 
    (normalized to HyGCN).}}
    \vspace{-0cm}
    \label{energy_saving}
  \end{minipage}
  \\
  \begin{minipage}[t]{1\linewidth}
    \centering
    \includegraphics[scale=0.5,trim=6.5cm 7.7cm 6.5cm 7.7cm,clip]{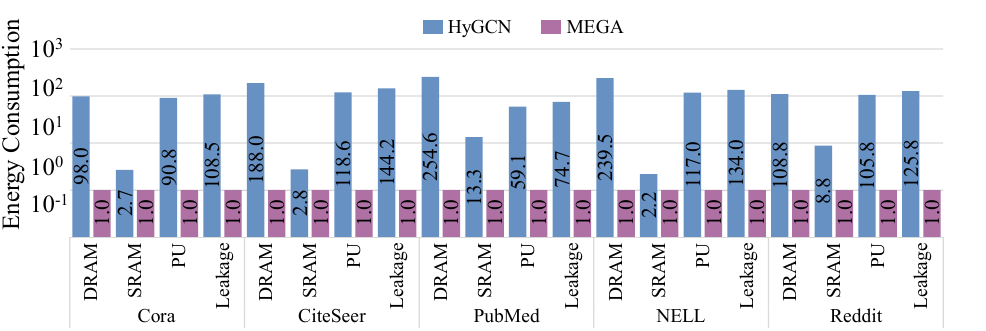}  
    \vspace{-0.2cm}
    \caption[short]{{Energy consumption breakdown comparisons with HyGCN on GCN 
    (normalized to our MEGA).}}
    \vspace{-0cm}
    \label{energy_breakdown_cmp}
  \end{minipage}
\end{figure}

\textit{\textbf{2) DRAM access:} }
As shown in Fig. \ref{dram_access},
on average, MEGA achieves $108.1\times$, $10.5\times$, 
$8.4\times$, and $7.3\times$ 
reduction on DRAM access 
over HyGCN, GCNAX, GROW, and SGCN, respectively.
The gains are mainly coming from two aspects:
(\romannumeral1) 
Our Degree-Aware 
method obtains a high compression ratio at the algorithm level.
(\romannumeral2) Adaptive-Package format 
transfers the  
theoretical memory benefits into practical DRAM access reduction to the maximum extent and 
the Condense-Edge approach significantly improves data reuse and further decreases DRAM access.
MEGA obtains significant DRAM access reduction over HyGCN because
HyGCN neither uses graph partition to improve data locality nor utilizes the feature sparsity.
While GROW consider the sparsity and adopt the partition algorithm to improve 
data locality,
there remains considerable irregular DRAM accesses
induced by the sparse connections,
which is substantially alleviated by our Condense-Edge approach in MEGA.

\textit{\textbf{3) Energy saving:}}
Fig. \ref{energy_saving} demonstrates that our accelerator has a significant 
energy-efficiency superiority 
compared to other state-of-the-art accelerators. On average, our MEGA achieves  
$47.6\times$, $7.2\times$, $5.4\times$, $4.5\times$ energy savings 
compared to HyGCN, GCNAX, GROW, 
and SGCN, respectively.
The energy efficiency gains are mainly attributed
to the significant reduction of DRAM access brought by our Degree-Aware 
method and the innovative Adaptive-Package format.
In addition, by supporting the mixed-precision computation,
the low-bit integers are used for computation instead of float-point data, 
further reducing energy consumption. 
Fig. \ref{energy_breakdown_cmp} displays a
breakdown analysis of the energy 
consumption in HyGCN and MEGA when running GCN.
The consumption is divided into four parts: DRAM, SRAM, 
Processing Unit (PU), and Leakage.
As the results show, MEGA has significant energy savings on all four parts, especially on 
DRAM access.

\subsection{Ablation Study}
To analyze where our gains come from and demonstrate 
the effectiveness of our three proposed techniques separately,
we conduct breakdown analysis on speedup and DRAM access.

\begin{figure}[t]
  \centering
    \includegraphics[scale=0.53,trim=0cm 0cm 0cm 0, clip]{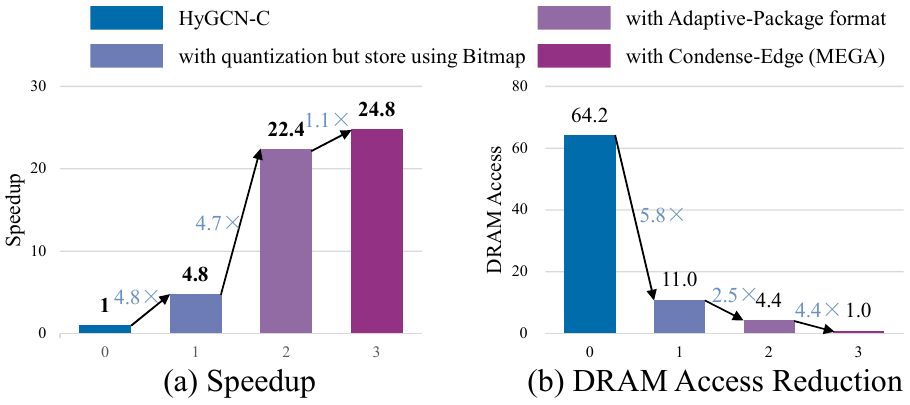}
    \vspace{-0.2cm}
      \caption{{(a) The contribution of each method to the improvement of performance.  
  (b) The contribution of each method to the DRAM access reduction.}}
    \vspace{-0.2cm}
    \label{speedup_dram_breakdown}
\end{figure}

\textit{\textbf{1) Speedup:}}
{Fig. \ref{speedup_dram_breakdown}(a) illustrates the contribution of each 
proposed method towards the speedup improvement of MEGA compared to HyGCN-C, which 
also adopts the execution
order of $\mathbf{A(XW)}$ to eliminate the contribution of the execution order 
in comparisons.}
When running the GCN quantized by our Degree-Aware method but 
in which the mixed-precision
features are stored in Bitmap format,
there is a significant performance improvement of
$4.8\times$ speedup on average.
Quantization not only decreases computational latency by using fixed-point operations, 
but also significantly decreases pipeline stall by reducing DRAM access.
When our Adaptive-Package sparse format is supported, 
MEGA obtains an average speedup of $4.7\times$ compared with the Bitmap format. The gain 
is mainly from that 
Adaptive-Package can store mixed-precision features with their own quantization bitwidths
rather than the highest bitwidth (8bit), 
which significantly reduces the latency of bit-serial computation and 
also decreases memory overhead.
Applying the Condense-Edge scheduling strategy provides a further $1.1\times$ speedup, 
in which  
the contribution 
is not significant because 
most cycles of the DRAM accesses can be overlapped, especially for quantized models.

\begin{figure}[t]
  \subfigure[Pipeline stall]{
    \begin{minipage}[t]{0.56\linewidth}
    \centering
    \includegraphics[scale=0.17,trim=0.2cm 0cm 0.2cm 0, clip]{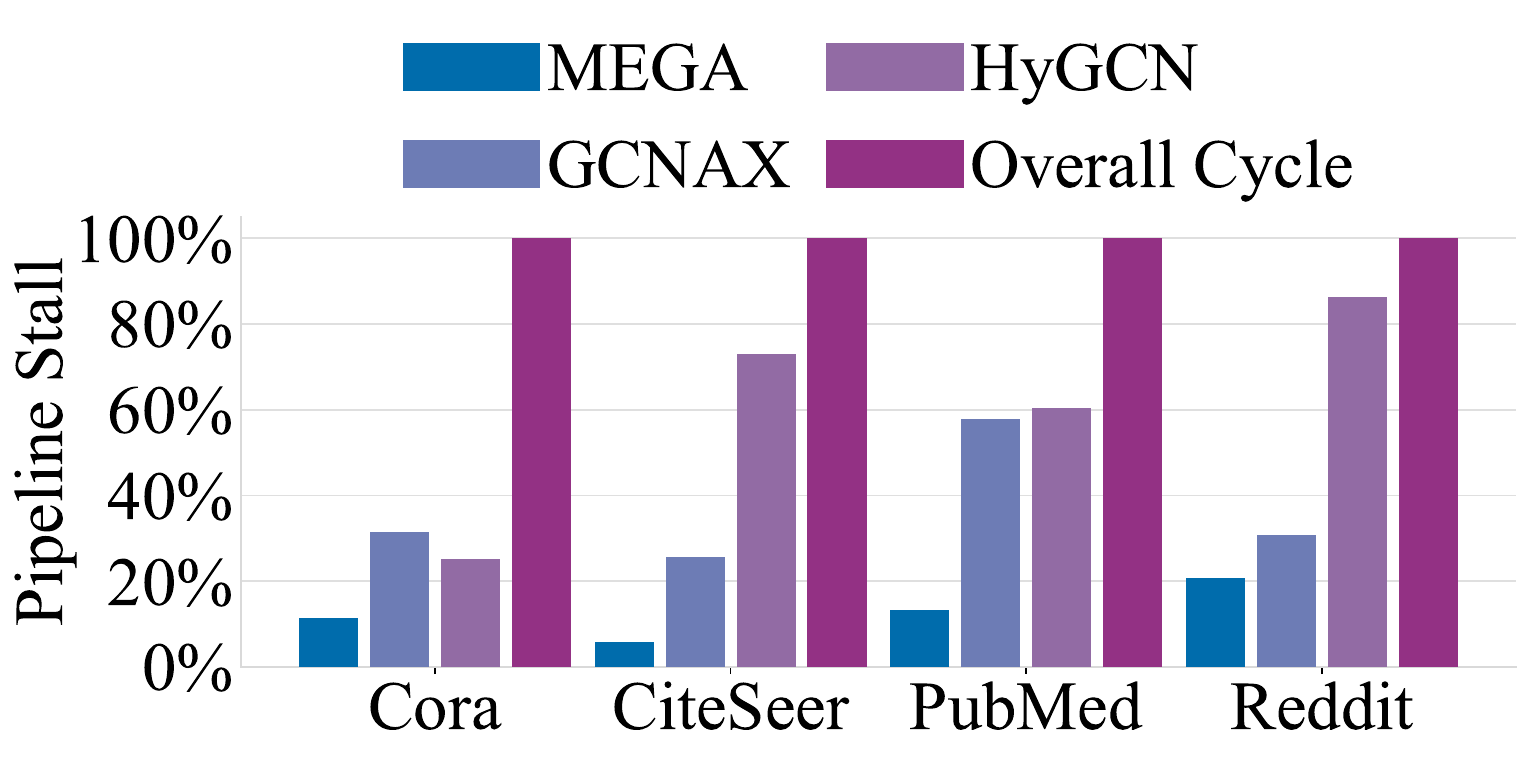}
    \vspace{-0.2cm}
    \vspace{-0.2cm}
    \label{dram_access_stall_cmp}
  \end{minipage}
  } 
  \subfigure[DRAM access]{
    \begin{minipage}[t]{0.36\linewidth}
      \centering
      \includegraphics[scale=0.17]{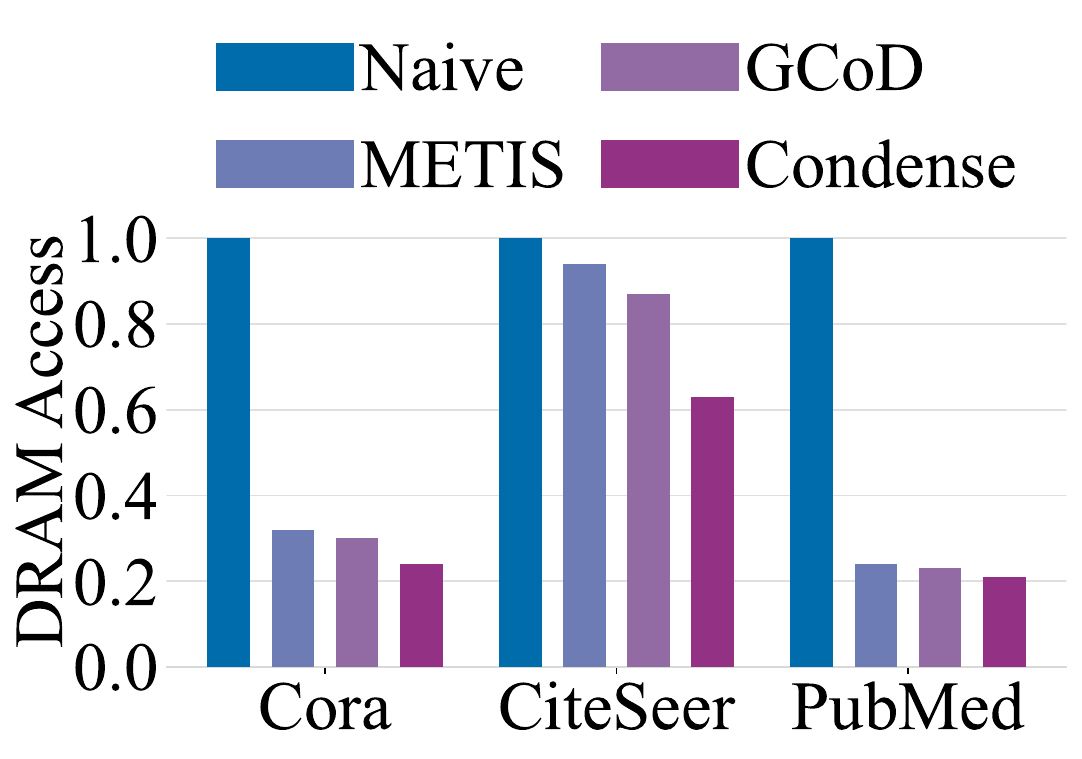}
      \vspace{-0.2cm}
      \label{condense_contribution}
      \vspace{-0.2cm}
    \end{minipage}
  }
  \vspace{-0.1cm}
  \caption{
  (a) The comparisons of stall induced by DRAM access.
  (b) The superiority of our Condense-Edge.
  }
  \vspace{-0cm}
  \label{analysis_on_stall}
\end{figure}

\textit{\textbf{2) DRAM access:}}
The breakdown analysis in Fig. \ref{speedup_dram_breakdown}(b)
demonstrates that our Degree-Aware quantization method 
is effective in decreasing DRAM access, which can be reduced by a factor of $5.8\times$ 
on average.
Furthermore, when supporting the Adaptive-Package format, MEGA reduces the DRAM access by
an average $2.5\times$.
Our Condense-Edge scheduling strategy also significantly contributes to 
reducing DRAM access by an average $4.4\times$, especially on the graph with a more 
sparse adjacency matrix, e.g., NELL (0.0073\%, $11.6\times$).
Based on our three proposed techniques, MEGA substantially reduces the stalls induced 
by the DRAM access compared to HyGCN and GCNAX, as shown in Fig. \ref{dram_access_stall_cmp}.

The DRAM access comparison between our Condense-Edge method and other methods of 
enhancing the data locality is shown in Fig. \ref{condense_contribution}.
`Naive' 
indicates that no method is used.
`METIS' denotes that divide the graph using METIS\cite{abou2006multilevel}.
GCoD partitions the graph using METIS and 
prunes the sparse connections. 
The 
comparison shows that 
our approach is superior to other methods in reducing DRAM access.

\subsection{Sensitivity Studies}
\label{sensitive studies}

\begin{figure}[t]
  \vspace{-0cm}
  \centering
    \includegraphics[scale=0.33]{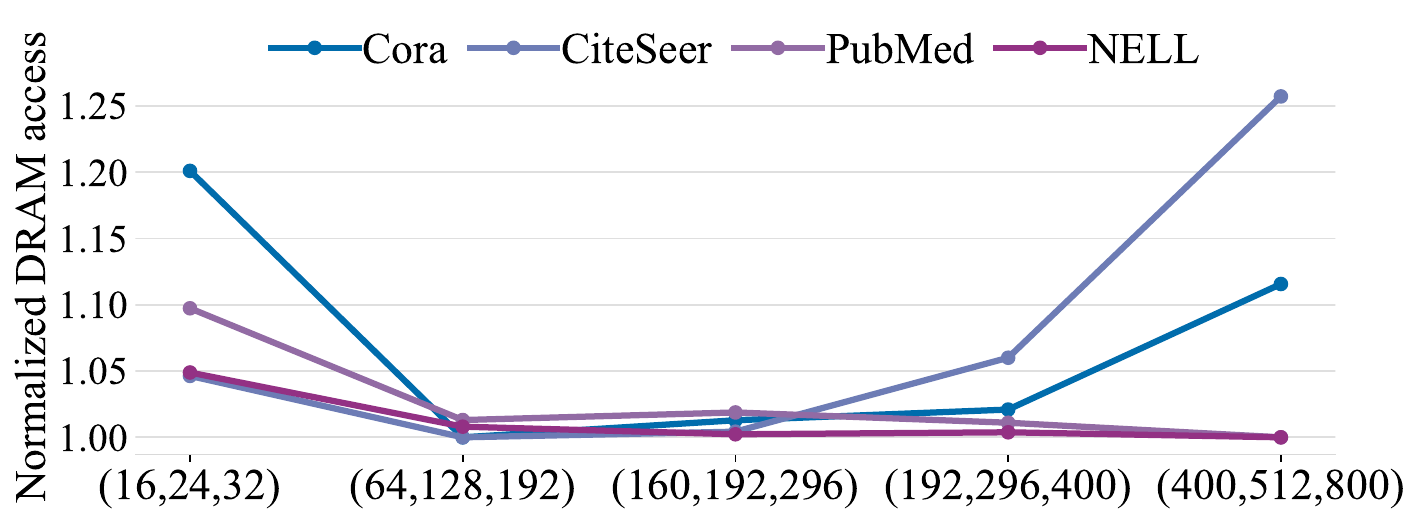}
    \vspace{-0.2cm}
    \caption{Design space exploration on the length settings of Adaptive-Package format
    (normalized to the optimal situation).
    }
    \vspace{-0cm}
    \label{val_length}
\end{figure}

\textit{\textbf{1) The length settings in Adaptive-Package format.}}
Since the three length settings in Adaptive-Package format are 
experimentally searched, we analyze the impacts of different 
settings on DRAM access, as shown in Fig. \ref{val_length}.
We observe that 
the settings to achieve the optimal case are various for different 
datasets.
For example, 
the optimal situation on PubMed is achieved by (400bits, 512bits, 800bits), in which 
the Cora and CiteSeer are much worse than their optimal cases. 
Considering the optimization across various datasets,
we finally adopt 
the setting of (64, 128, 192) in our design.

\begin{figure}[t]
  \vspace{-0cm}
  \centering
    \includegraphics[scale=0.38]{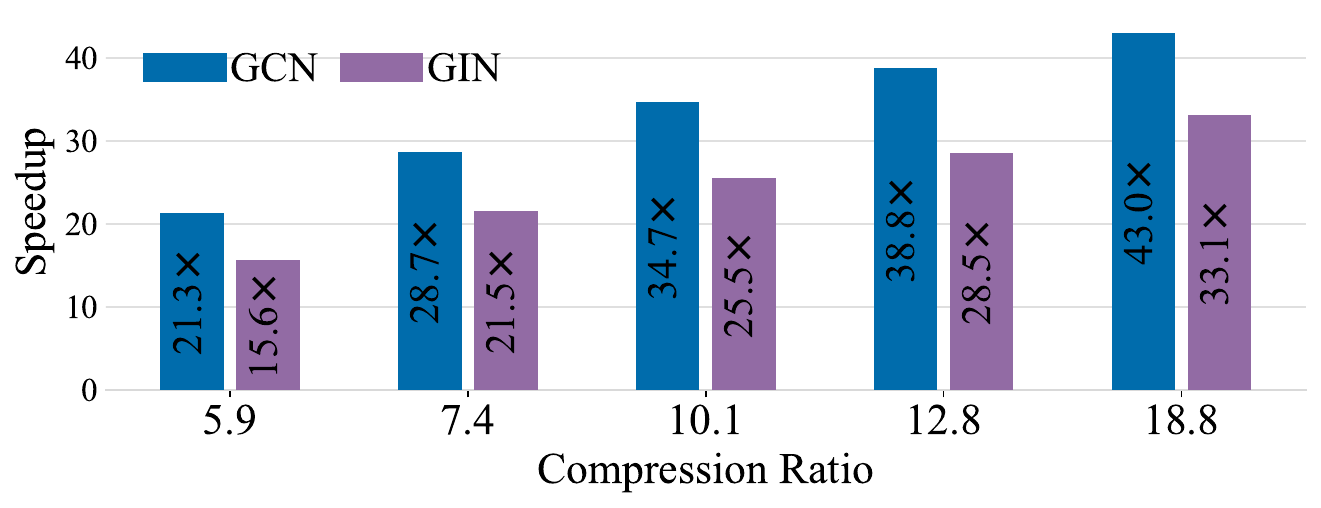}
    \vspace{-0.2cm}
    \caption{{Sensitivity to the compression ratios on Cora of GCN 
    and GIN (normalized to HyGCN).}}
    \vspace{-0cm}
    \label{abcd}
\end{figure}

{\textit{\textbf{2) Sensitivity on the compression ratio.}}
As an accelerator that supports the storage and computation of mixed-precision node features, 
the compression ratio may significantly affect MEGA's performance. Therefore, 
as illustrated in Fig. \ref{abcd},
we explore the sensitivity of MEGA's performance to the compression ratio. 
We can observe that the performance of 
MEGA scales well with the increased compression ratio.}

\section{Discussion}
{\textit{\textbf{1) The training overhead of Degree-Aware quantization method.}}
We compare training time between FP32 models and quantized models on various datasets and 
GNN models. On average, the training time of 
quantized models is 2.04x over the FP32 models and the memory overhead 
is up to 12.8\% on one 3090GPU. The training cost in both time and memory is acceptable.
The training overhead of our method is much less than the state-of-the-art GNN quantization 
method DQ, which 
on-average costs 1.58x training time compared with ours.}

\textit{\textbf{2) The broader applicability of the Condense-Edge.}}
The Condense-Edge method is initialized by partitioning graphs using METIS, which may
become inefficient when processing dynamic or 
large-scale graphs. However, in such cases, our Condense-Edge method can work without 
partitioning while reaping comparable gains in speedup and energy efficiency. 
For instance, compared with 
the state-of-the-art accelerator SGCN, our MEGA without partitioning on-average obtains 
$3.50\times$ ($\downarrow $3\% compared to MEGA using METIS) 
and $3.95\times$ ($\downarrow $14\%) improvement 
on speedup and energy efficiency, respectively. 
{This indicates that 
our Condense-Edge method works well 
without partitioning (just with a little performance discount) 
because the connections in the original graph are also extremely sparse, 
which is a challenge effectively addressed by the reorder process in our 
Condense-Edge method.}
Therefore, our Condense-Edge can be applied widely to various types of graphs, 
including dynamic or large-scale graphs.

\textit{\textbf{3) The support of Graph Attention Network (GAT).}}
With the same combination phase as GCN, 
GAT\cite{velivckovic2018graph} 
employs the attention mechanism as its 
aggregation function,
which involves MLP and softmax operators. 
Quantized by our Degree-Aware method, 
GAT obtains up to $16.5\times$ (from $9.6\times$) compression ratio  
with negligible accuracy loss (e.g., on CiteSeer, FP32: 72.5\%, Degree-Aware: 71.9\%). 
MLP can
be computed reusing MEGA's Combination Engine and there are 
multiple prior literature\cite{ham20203,marchisio2022enabling,stevens2021softermax} 
on hardware-accelerated softmax, which can be directly integrated into MEGA with 
minor changes to the dataflow.
For example, supporting
GAT is estimated to incur a 1.5\% area overhead using the 
hardware-accelerated softmax design in $A^3$\cite{ham20203}.

\section{Conclusion}
In this paper, we propose an algorithm and hardware co-design framework to boost 
the memory efficiency
in GNN acceleration. Specifically, at the algorithm level, we
propose the 
Degree-Aware mixed-precision quantization method, 
in 
which a proper bitwidth is learned and
allocated to a node according to its in-degree to compress
GNNs as much as possible while retaining accuracy. 
At the hardware level, we propose our MEGA accelerator adopting 
an innovative sparse format named Adaptive-Package to 
thoroughly transfer the theoretical 
gains on memory
into an empirical improvement of performance and energy-efficiency.
We also present a 
Condense-Edge scheduling strategy to alleviate the
numerous and irregular DRAM accesses caused by the 
extremely sparse adjacency matrix.
Extensive experiments at the algorithm and hardware levels demonstrate the superiority 
of our MEGA.

\section*{Acknowledgements}
This work was supported in part by the STI 2030-Major Projects (No.2021ZD0201504), 
the National Natural Science Foundation of China (No.62106267, NO.61972242), 
the Jiangsu Key Research and Development Plan (No.BE2023016, No.BE2021012-2).


\bibliographystyle{IEEEtranS}
\bibliography{refs}

\end{document}